\pdfoutput=1 
\documentclass[a4paper,11pt]{article}
\usepackage{jheppub} 
\usepackage[T1]{fontenc} 
\usepackage{mathtools}
\usepackage{physics}
\usepackage[utf8]{inputenc}
\usepackage[normalem]{ulem}
\usepackage[dvipsnames]{xcolor}
\usepackage{mathrsfs}
\usepackage{mathtools}
\usepackage{mathrsfs}
\usepackage{footnote}
\usepackage{fontenc}
\usepackage{wasysym}
\usepackage{physics}
\usepackage{tabularx}
\usepackage{lscape}
\usepackage[makeroom]{cancel}
\usepackage[marginal,stable,bottom]{footmisc} 
\renewcommand{\bar}{\overline}
\newcommand{\E}{\mathcal{E}}
\newcommand{\X}{X}
\newcommand{\Xh}{\hat{\X}}
\newcommand{\XX[1]}{\X_{#1}}
\newcommand{\Y}{Y}

\newcommand{\I}{{\smallint}}
\newcommand{\F}{F}

\newcommand{\Z}{{Z}}

\newcommand{\Rh}{\hat{\mathrm{R}}}

\newcommand{\Rm}{\mathcal{R}}
\newcommand{\Lm}{\mathcal{L}}
\newcommand{\Lp}{\mathcal{L}^{\prime}}

\newcommand{\Lmh}{\hat{\mathcal{L}}}
\newcommand{\LG}{\mathrm{G}}

\newcommand{\GGT}{\mathrm{F}}

\newcommand{\GG}{G}
\newcommand{\Gh}{\hat{G}}
\newcommand{\MG}{M}
\newcommand{\LGh}{\hat{\LG}}
\newcommand{\gh}{\hat{g}}
\newcommand{\phih}{\hat{\phi}}

\newcommand{\xh}{\hat{x}}
\newcommand{\bx}{\psi}
\usepackage{bm}
\DeclareMathAlphabet{\mathpzc}{OT1}{pzc}{m}{it} 
\DeclareSymbolFont{anttfont}{OML}{antt}{m}{it}

\DeclareMathSymbol{\ff}{\mathalpha}{anttfont}{`f}

\usepackage{autobreak} 
\usepackage[colorinlistoftodos]{todonotes}
\presetkeys{todonotes}{disable}{} 

\newcommand{\add}[1]{\textcolor{blue}{#1}}

\definecolor{rosy}{RGB}{230,235,252}
\definecolor{myframetitle}{RGB}{90,89,170}
\definecolor{myblocktitle}{RGB}{140,185,249}
\definecolor{mytitle}{RGB}{10,80,26}

\definecolor{darkgreen}{RGB}{27,130,45}
\definecolor{darkblue}{rgb}{0,0,0.3}
\definecolor{darkred}{rgb}{0.7,0,0}

\definecolor{light gray}{RGB}{220,220,220}
\definecolor{dark purple}{RGB}{108,0,217}
\definecolor{pink}{RGB}{190,20,100}
\definecolor{orang}{RGB}{193,63,0}
\definecolor{green}{RGB}{11,98,17}
\definecolor{darkpink}{RGB}{153,0,76}
\definecolor{bluegreen}{RGB}{0,102,102}
\definecolor{greenlagan}{RGB}{0,102,0}
\definecolor{redgreen}{RGB}{102,102,0}
\definecolor{Redgreen}{RGB}{153,76,0}
\definecolor{vividviolet}{rgb}{0.62, 0.0, 1.0}
\definecolor{amaranth}{rgb}{0.9, 0.17, 0.31}
\definecolor{palatinateblue}{rgb}{0.15, 0.23, 0.89}
\definecolor{brightpink}{rgb}{1.0, 0.0, 0.5}
\definecolor{cornflowerblue}{rgb}{0.39, 0.58, 0.93}
\definecolor{deepcarminepink}{rgb}{0.94, 0.19, 0.22}
\definecolor{radicalred}{rgb}{1.0, 0.21, 0.37}

\usepackage[most]{tcolorbox}
\newcommand\hlbox[1]{\tcbhighmath{#1}}
\tcbset{highlight math style={left=02mm,right=02mm,top=02mm,bottom=02mm}} 
\hypersetup{ linktoc=all,
    colorlinks, linkcolor={palatinateblue},
    citecolor={brightpink}, urlcolor={amaranth}
}

  \makeatletter  
\gdef\@fpheader{}  
\makeatother 
\begin{document}


\title{ \Large \centering Two-dimensional (bi-)scalar gravities \\ from four-dimensional Horndeski}

\author[\dagger, \ast]{M. Shams Nejati,}
\author[\dagger,\star]{M.H.~Vahidinia}

\affiliation[\dagger]{Department of Physics, Institute for Advanced Studies in Basic Sciences (IASBS),
P.O. Box 45137-66731, Zanjan, Iran}
\affiliation[\star]{School of Physics, Institute for Research in Fundamental
Sciences (IPM),\\ P.O.Box 19395-5531, Tehran, Iran}

\affiliation[\ast]{Institute for Theoretical Physics, TU Wien, Wiedner Hauptstrasse 8–10/136, A-1040 Vienna, Austria}

\emailAdd{m\textunderscore shams@iasbs.ac.ir,
vahidinia@iasbs.ac.ir}

\abstract{
We develop a classical two-dimensional bi-scalar gravity based on the Kaluza-Klein reduction applied to the four-dimensional Horndeski theory. One of the scalar fields arises from the original four-dimensional theory, while the extra scalar emerges from the reduction process. We also introduce a two-dimensional bi-scalar identity that allows for a more concise and elegant reformulation of the resulting bi-scalar Lagrangian. Additionally,  we study the linear perturbations around a static background to demonstrate that the bi-scalar theory may support a single healthy propagating mode. Furthermore, by restricting the scalar fields, we investigate a general single scalar theory that is identical to the two-dimensional Horndeski theory up to a boundary term. Our results provide a framework to map a generic two-dimensional dilaton gravity into four-dimensional Horndeski theory.}
\maketitle
\section{Introduction}
Theories of gravity in lower dimensions are essential for understanding gravitational physics, including issues related to black holes and other related topics. Lowering space-time dimensions not only simplifies the computation but also improves the knowledge. Furthermore, specific problems regarding semi-classical and quantum aspects of black holes, such as information paradox, black hole microstates, etc., may be independent of space-time dimensions and are better explored in lower dimensions. Moreover, some systems exhibit an effective space-time dimension lower than the actual dimension. For instance, the near horizon of extremal black holes can be investigated as an effectively two-dimensional or three-dimensional system (see e.g. \cite{Kunduri:2013gce, Sadeghian:2015laa, Sadeghian:2015nfi}). In addition, the lower-dimensional theory can be employed to model the dynamics of spherically symmetric solutions in higher dimensions \cite{Kunstatter:2015vxa}. 
	
Among lower-dimensional theories, 2D is significant because it represents the lowest dimension where the notion of space-time remains relevant. The pure 2D Einstein theory is topological, limiting its application to specific scenarios. To allow some degrees of freedom (at least on the boundary), we need to modify it by including extra fields. The most straightforward approach is adding a scalar degree of freedom to the system, usually called dilaton. Dilaton gravity can also arise in other ways, like investigating non-critical string theory or as a toy model for various aspects of classical and quantum black holes (see e.g. \cite{Grumiller:2002nm, Nojiri:2000ja}). Jackiw-Teitelboim (JT) gravity \cite{Jackiw:1984je, Teitelboim:1983ux} as an important dilaton theory, has taught us a lot in different contexts (see, e.g. \cite{Mertens:2022irh} for a recent review). Interestingly, the JT gravity is regarded as the holographic dual of the Sachdev-Ye-Kitaev model \cite{Kitaev:2015, Maldacena:2016hyu}. This duality 
offers a novel setting to investigate the connections between gravity and quantum field theories. Additionally, JT gravity is also related to matrix models, which are used to study non-perturbative aspects of quantum gravity \cite{Saad:2019lba}.

In addition to JT gravity, various other 2D gravitational theories, including its generalization, have been explored in literature. In particular, Grumiller, Ruzziconi, and Zwikel (GRZ) have investigated the most general consistent deformation of JT gravity that preserves Lorentz invariance \cite{Grumiller:2021cwg}.  This theory encompasses various 2D dilaton theories such as Callan-Giddings-Harvey-Strominger (CGHS) \cite{Callan:1992rs, Strominger:1994tn}, and models that are explored in \cite{Almheiri:2014cka}.On the other hand, we argued in \cite{Nejati:2023hpe} that the most general 2D scalar-tensor gravity with second-order field equations can be obtained by applying a disformal transformation to JT gravity. As one might expect, the resulting theory (up to a boundary term) belongs to the Horndeski theory. This theory represents the most general scalar-tensor gravity with second-order field equations in dimensions $D=2$, $D=3$, and $D=4$, as it has been classified \cite{Horndeski:1974wa}.

Nevertheless, it is important to note that the mentioned 2D dilaton theories of gravity do not have propagating degrees of freedom. Introducing dynamics into the theory can be achieved by adding an extra scalar field, leading to constructing a bi-scalar gravitational theory. Indeed, a primary motivation for this study is to explore the most general two-dimensional bi-scalar theory. One way to obtain this demand is by utilizing Kaluza-Klein (KK) dimensional reduction over a suitable ansatz, which is aligned with a symmetry existing in the system (e.g. spherical symmetry). In the absence of such symmetry, or in the presence of higher derivative terms, the resulting theory may have different dynamics to its higher-dimensional origin \cite{Nojiri:1999br,Katanaev:1998ry}. This approach allows us to obtain dilaton gravity from higher-dimensional Einstein theory, where the dilaton field is associated with the radius of compactification. In particular, JT gravity can be embedded in higher-dimensional Einstein-Maxwell-Dilaton \cite{Li:2018omr} and 2D two-dilaton theory (which is a simple bi-scalar theory, linear in the kinetic terms) into 4D Brans-Dicke theory \cite{Grumiller:2000wt}. Moreover, the 4D Horndeski theory can also be obtained by dimensionally reducing the Lovelock theory. \cite{VanAcoleyen:2011mj}. Here, we will follow a similar approach, starting from the 4D Horndeski theory (considered the most general second-order single-scalar theory) to obtain a general 2D bi-scalar theory. In this case, one of the scalar fields emerges from the original 4D Horndeski theory, and the other one is a dilaton field --- a consequence of KK reduction. 

The organization of this paper is as follows. In Sec. \ref{sec:KK}, we obtain a bi-scalar theory by reducing the dimensions of a 4D Horndeski theory on a 2D sphere and attain a 2D bi-scalar theory. In Sec. \ref{sec:bi-scalar alternative}, we introduce an identity, that allows us to simplify the form of 2D bi-scalar Lagrangian up to a total derivative (that does not impact field equations). The new simplified version of the bi-scalar theory suggests a more comprehensive form of the bi-scalar Lagrangian. In Sec. \ref{sec:bi-scalar solutions}, we present some examples to show how solutions to 4D Horndeski map to solutions of bi-scalar theory in two dimensions. Afterwards, in Sec. \ref{sec:pert} we perform linear perturbations around a static background to show that bi-scalar theory may support a healthy propagating mode. In Sec. \ref{sec:signle-scalar} we impose a constraint on the bi-scalar theory to obtain a generic single-scalar Lagrangian that is indeed a 2D version of Horndeski theory.  This theory includes 2D dilaton theories such as JT, GRZ, and CGHS. In addition, we classify all possible embedding of these 2D theories in 4D Horndeski. Besides, we also explicitly verify the consistency of the field equation that arises from reduced action.  Finally, Sec. \ref{sec:Remarks} is devoted to a summary of results, some remarks, as well as possible future directions.

\textbf{Note:} Upon completion of our paper, we became aware of \cite{Mandal:2023kpu} that partially overlaps with section \ref{sec:KK} of our work on Kaluza-Klein reduction in a specific 4D Horndeski theory.

\textbf{Notations and conventions.} In what follows, $c,d,...={0,1,2,3}$ are indices of four-dimensional spacetime, $\mu,\nu,...={0,1}$ are indices of two-dimensional spacetime  and  $i,j,...={2,3}$ are related to extra dimensions of 2D sphere. Besides, quantities denoted by ``hat''  are 4D quantities, while the simple quantities refer to 2D ones. We also use $I,J, \cdots = 1,2$ to denote scalar fields  $\phi^{I}$ as $\phi:=\phi^{1}$ and $\psi:=\phi^{2}$. In this regards, $\XX[IJ]:=-\frac{1}{2} \nabla_\mu \phi^I \nabla^\mu \phi^J$ accordingly. Finally, we find it convenient to use $\X=\XX[11]$, ${\Y}=\XX[22]$, and ${\Z}=\XX[12]$ in certain contexts.


\section{Bi-scalar theory from KK reduction}\label{sec:KK}
The most general Lagrangian which is made of metric and a scalar field in four dimensions and leads to second-order equations of motion  is Horndeski theory with the action \cite{Horndeski:1974wa,Deffayet:2011gz}:
\begin{gather}
    S=\int \dd[4]x \sqrt{\abs{\gh}} \, \sum^{5}_{\mathrm{n}=2} \Lmh_n \; ,
\end{gather}
in which
\begin{eqnarray} \label{Horndeski}
	\Lmh_{2} &=& G_{2} (\phih , \Xh) \; , \qquad
	\mathcal{L}_{3} = G_{3} (\phih , \Xh) \, \hat{\Box} \phih \; , \nonumber\\
	\Lmh_{4} &=& G_{4} (\phih , \Xh)  \, \Rh +  G_{4,\Xh}(\phih , \Xh)  \qty((\hat{\Box} \phih)^{2}-(\nabla_{c}\nabla_{d}\phih)^2)\; ,\\
	\Lmh_{5} &=& G_{5}(\phih, \Xh) \, \LGh_{cd} \nabla^{c} \nabla^{d} \phih - \frac{1}{6}  G_{5,\Xh}
	(\phih , \Xh) \qty( (\hat{\Box} \phih)^{3} -3 \hat{\Box} \phih (\nabla_{c} \nabla_{d} \phih)^2+ 2 (\nabla_{c} \nabla_{d} \phih)^3 ),  \nonumber
\end{eqnarray}
where $G_{n}$ is an arbitrary function of scalar field $\hat{\phi}$ and its kinetic term $\Xh:=-\frac{1}{2}\partial_{c}\phi \partial^{c}\phi$ while $G_{n,\X}:=\partial_\X G_n$. Note that $\Rh$ represents the 4D Ricci scalar, while $\LGh_{cd}$ denotes the Einstein tensor in four dimensions, and  the powers of second derivatives of scalar fields are defined as $(\nabla_{c} \nabla_{d} \phih)^2= \nabla_{c} \nabla_{d} \phih \nabla^{c} \nabla^{d} \phih$ and
$(\nabla_{c} \nabla_{d} \phih)^3= \nabla_{c} \nabla_{d} \phih \nabla_{e} \nabla^{c} \phih \nabla^{d} \nabla^{e} \phih$.

As mentioned in the introduction, we are aiming to build a 2D bi-scalar theory, using the KK dimensional reduction. For this purpose, we choose the metric ansatz as 
\begin{eqnarray} \label{eq:reduction-ansatz}
  d\hat{s}^{2}&=&\gh_{cd} \; d\xh^{c} d\xh^{d}= g_{\mu \nu}(x) \; dx^{\mu} dx^{\nu}+\bx^{2}(x) \gamma_{i j}(\theta) \; d\theta^{i} d\theta^{j} ,
\end{eqnarray}
in which $\gamma_{i j}(\theta)$ is the 2D sphere metric. 
 Spherical symmetry implies that the scalar field does not depend on angular coordinates $\theta$, hence
\begin{eqnarray}
  \phih(\xh)=\phi(x) \rightarrow \Xh(\xh)=\X(x).
\end{eqnarray}
Under this reduction, 4D derivatives of the scalar field and 2D ones, are related (for more details see the appendix \ref{app:KK})
\begin{equation}
    \begin{aligned}
  \hat{\Box} \phi &=\Box \phi - 4 \bx^{-1}\,\XX[12] \,\\
 (\nabla_{c}\nabla_{d}\phi)^ 2& =(\nabla_{\mu}\nabla_{\nu}\phi)^ 2+8 \bx^{-2}\,\XX[12]^2\; ,\\
  (\nabla_{c}\nabla_{d}\phi)^3&=(\nabla_{\mu}\nabla_{\nu}\phi)^3-16 \bx^{-3}\, \XX[12]^{3},
\end{aligned}
\end{equation}
where $\XX[IJ]:=-\frac{1}{2} \nabla_\mu \phi^I \nabla^\mu \phi^J$ and $\{ \phi^1:= \phi,\, \phi^2:=\psi \}$.
In addition, the 4D Ricci scalar relates to the 2D one 
\begin{gather}
    \Rh=R+ \bx^{-2} \Rm-4 \bx^{-1} \Box \bx -2 \bx^{-2}\nabla_{\mu}\bx \nabla^{\mu}\bx \; ,
\end{gather} 
in which $R$ is 2D Ricci scalar and $\Rm$ is the curvature of $\gamma_{ij}$, being $\Rm=2$ for a unit 2D sphere. Finally, the spherical symmetry implies that Einstein tensor 
lacks a vector part $\LGh_{\mu i}=0$,  and the remaining components are given by
\begin{equation}
    \begin{gathered}
  \LGh_{\mu \nu}=-\frac{2}{\bx}\nabla_{\mu}\nabla_{\nu}\bx-\frac{1}{2} g_{\mu \nu} \qty(\Rh-R), \\
  \LGh_{ij}=\gamma_{ij}\qty(\bx\Box \bx-\frac{\bx^2}{2} R)\; .
\end{gathered}
\end{equation}
Consequently, by noting $\sqrt{\abs{\gh}}=\psi^2 \sqrt{\abs{\gamma}}\sqrt{\abs{g}}$ we obtain the 2D bi-scalar action from KK reduction of 4D Horndeski theory
\begin{gather}
    S=4\pi \int \dd[2]x \sqrt{\abs{g}} \Lm \; ,
\end{gather}
with the following Lagrangian density:
\begin{gather}
\label{eq:2Dbiscalar}
\hlbox{\mathcal{L} = F_2+F_3^{I} \Box \phi^{I}+F_4 R +F_{4, \XX[1I]} L_{1I}.}
\end{gather}
Here, summation over repeated index $I$ is assumed, and
 $L_{IJ}$ is defined as
\begin{equation}
    L_{IJ}:=\Box \phi^I \Box \phi^J -\nabla_\mu \nabla_\nu \phi^I \nabla^\mu \nabla^\nu \phi^J.
\end{equation}
Note that the  scalar functions $F_{n}(\phi^{I},\X_{IJ})$ are related to $\GG_n$ which can be rewritten as
\begin{align} 
    F_2&=\bx^2 G_2-4\bx \XX[12]G_3+2(1+2\XX[22])G_4 +8\XX[12]^2 G_{4,\XX[11]} ,\nonumber\\
F_3^1&=\bx^2 G_3 -8\bx \XX[12] G_{4,\XX[11]}-G_5(1+2\XX[22])-4\XX[12]^2 G_{5,\XX[11]} , \\
F_3^2&=-4 \qty( \bx G_4 +\XX[12] G_5), \quad
 F_4 =\psi^2 G_4+2 \psi G_5 \XX[12] .\nonumber
\end{align}
It's important to note that the KK reduction is an on-shell procedure, and therefore, it should be applied to equations of motion. Therefore, verifying the consistency of the field equations derived from the reduced Lagrangian is necessary. 
The initial step is to verify if the reduced Lagrangian \eqref{eq:2Dbiscalar} allows for second-order field equations.
Considering the resulting Lagrangian is in the form of the standard bi-scalar theory \eqref{4DBiscalar}, it guarantees that it provides second-order field equations \cite{Ohashi:2015fma}, as suggested by the original 4D Lagrangian. However, in Sec. \ref{sec:field-eq}, we will explicitly examine the consistency of field equations in particular situations.

\section{An alternative form of the bi-scalar Lagrangian}\label{sec:bi-scalar alternative}
In the preceding section, we have found the general bi-scalar theory that arises from 4D Horndeski. 
Here, we will show how a certain 2D identity enables us to express it in an alternative form.

The main observation comes from the following simple identity between Ricci scalar and derivative of the scalar field \footnote{
To derive this identity, we can begin with the definition of the Riemann tensor 
 $\comm{\nabla_\mu}{\nabla_\nu} \nabla^\rho \phi= R^{\rho}_{\; \sigma \mu \nu} \nabla^\sigma \phi$ and contract indices $\rho$ and $\mu$ and contract two sides with $\nabla^{\nu}\phi$. Then  using the 
 $R_{\mu \nu}=\frac{1}{2} R g_{\mu \nu}$ in two-dimensions we can obtain the desired result. 
}
\begin{gather} \label{eq:2D-identiy-L4}
    \X R +L_{11}
  =\nabla_\mu (\nabla^\mu \X+ \Box \phi \nabla^\mu \phi).
\end{gather}
This identity suggests that one may eliminate 
$F_4 (\XX[11]) R+F_{4,X_{11}}L_{11}$ from the action up to a boundary term as happens in the 2D Horndeski  \eqref{Lag:2D Horndeski}  \cite{Takahashi:2018yzc,Nejati:2023hpe}.  However, one may note that the extra  scalar field  $\psi$  also has contributions to the families, and hence
 \eqref{eq:2Dbiscalar} also incorporates the term $L_{12}$, which requires attention. Hopefully, it is possible to generalize the mentioned identity for  functions of $\XX[IJ]$ and $\phi^{I}$  (see appendix \ref{app:identity-bi-scalar})
\begin{equation}\label{eq:2D-identiy-L4-main}
  \hlbox{ M R+ M_{,\XX[IJ]} L_{IJ}  = \nabla_\mu K^\mu+f_{3}^{I} \Box \phi^{I}+f_2 }
\end{equation}
as far as $ M_{,\XX[12],\XX[12]}=4 M_{,\XX[11],\XX[22]}$ holds for $M=M(\phi,\psi,\XX[11],\XX[12],\XX[22])$.
The details of the identity and the functions 
$K^\mu$, $f_{3}^{I}$, and $f_2$
for  $M=\XX[12] G(\XX[11])+H(\XX[11])$  that is relevant to our case ($M=F_4$) are given in \eqref{function-identity}.  Consequently, one can employ this identity for $M=F_4$ to simplify Lagrangian of bi-scalar theory to 
\begin{gather}\label{eq:2D-KGB-like}
    \mathcal{L} =(F_2+f_2)+(F_3^\phi+f_3^\phi) \Box \phi+(F_3^\psi+f_3^\psi)  \Box \psi+\nabla_\mu K^\mu .
\end{gather}
Therefore, by defining 
$\GGT_{n}:=f_n+F_{n}$ and up to a boundary term,
there is an alternative form for the bi-scalar Lagrangian \eqref{eq:2Dbiscalar}
\begin{gather}\label{eq:2D-KGB-like-2}
    \hlbox{\mathcal{L} =\GGT_2+\GGT_3^{I} \Box \phi^{I},}
\end{gather}
that results in a second-order field equation for the metric  $\E^{\mu \nu}=0$ where
    \begin{align}
        \E^{\mu \nu} &= \qty(\GGT_{2}+2\GGT^{(I}_{3,\phi^{J})} \XX[IJ]) g^{\mu \nu}+\qty(\GGT_{2,IJ}+2\GGT^{(I}_{3,\phi^{J})} ) \nabla^{\mu} \phi^{I} \nabla^{\nu} \phi^{J} \\
        &-2\XX[JK] \GGT_{3,JK} g^{\gamma (\mu} \delta^{\nu) \rho}_{\gamma \sigma} \, \nabla^{\sigma}\nabla_{\rho} \phi^{I}-\GGT_{3,JK}\; g^{\gamma (\mu} \delta^{\nu) \rho \zeta}_{\gamma \sigma \eta} \, \nabla_{\rho} \phi^{I} \nabla^{\sigma} \phi^{J}\nabla^{\eta} \nabla_{\zeta} \phi^{K},\nonumber
    \end{align}
 in which $F_{,IJ}:= \frac{1}{2}\qty(F_{,\XX[IJ]}+F_{,\XX[JI]})$
and $\GGT^{(I}_{3,\phi^{J})}:=\frac{1}{2}(\GGT^{I}_{3,\phi^{J}}+\GGT^{J}_{3,\phi^{I}})$ . Additionally, the scalar fields $\E^{I}=0$ where 
    \begin{align}
        \E^{I} &=\GGT_{2,\phi^{I}}-2 \XX[JK] \GGT_{2,\phi^{K},IJ}+\GGT_{2,IJ,KL}\nabla_\mu \XX[KL]\nabla^\mu \phi^{J}+ \GGT_{2,IJ} \Box \phi^{J}
         +\GGT^{J}_{3,\phi^{I}} \Box \phi^{J} \nonumber\\
         &-2 \GGT^{I}_{3,\phi^{J},\phi^{K}}\XX[JK]+ \GGT^{I}_{3,JK,LM} \nabla_\mu \XX[JK] \nabla^\mu \XX[LM]+\GGT^{I}_{3,JK}(\XX[IJ] R+L_{JK})\\
       &-2 \GGT^{J}_{3,KI,\phi^{L}} \XX[KL] \Box \phi^{J} +2\GGT^{I}_{3,JK,\phi^{L}}\nabla_{\mu} \XX[JK] \nabla^{\mu} \phi^{L}+ \GGT^{J}_{3,IK,LM} \nabla_{\mu} \XX[LM] \nabla^{\mu} \phi^{K} \Box \phi^{J} \nonumber
    \end{align}
Note that these equations of motion involve the contribution of $F_4$ terms through the $\phi^{I}$ dependence of functions $f_2$ and $f^{I}_3$.

The Lagrangian \eqref{eq:2D-KGB-like-2} is one of the main results of this paper. It's important to highlight that the Lagrangian functions satisfy two constraints: $\GGT^{\psi}_{3,\XX[12]}=2\GGT^{\phi}_{3,\XX[22]}$ and $\GGT^{\phi}_{3,\XX[12]}=2\GGT^{\psi}_{3,\XX[11]}$\add{\footnote{Using the notation $F_{,IJ}:= \frac{1}{2}\qty(F_{,\XX[IJ]}+F_{,\XX[JI]})$, one can represent the constraints as $F^{I}_{,JK}=F^{J}_{,IK}$.}}. These constraints were crucial for having second-order field equations. Thus, even though the components of the Lagrangian have a specific form in $\XX[12]$ and $\XX[22]$, one can promote the functions to arbitrary functions $\GGT_n=\GGT_n(\phi^{I}, \XX[IJ])$  which satisfy these constraints and find a more general 2D bi-scalar-tensor theory that leads to second order field equations. 

Indeed, the Lagrangian \eqref{eq:2D-KGB-like-2} has the reminiscent of some families of 4D bi-scalar theory as presented by Ohashi et al. in \cite{Ohashi:2015fma} as a candidate for the most general bi-scalar theory leading to second order field equations (see also \cite{Nicolis:2008in,Deffayet:2009wt,Deffayet:2009mn,Padilla:2010de,Padilla:2012dx,Sivanesan:2013tba,Kobayashi:2013ina,Charmousis:2014zaa}).
 However, more recently, Horndeski investigated the same problem and showed that the second-order equations of motion are more restricted than what was performed in  \cite{Ohashi:2015fma}. He proves that the Lagrangian corresponding to these second-order equations of motion can be found by $\Lm=g^{cd} \E_{cd}$, and finally provides the most general 4D bi-scalar Lagrangian \cite{Horndeski:2024hee}.
 
 For further investigation, we have considered the Lagrangian provided by Ohashi et al. \cite{Ohashi:2015fma} as the correct bi-scalar Lagrangian in two dimensions. We then applied 2D identities to obtain \eqref{eq:2D-KGB-like-2} (see the appendix \ref{app:4D-bi-scalar} for details).

\section{2D bi-scalar solutions from 4D Horndeski }\label{sec:bi-scalar solutions}
Through the process of dimensional reduction, one can establish a relationship between the solutions of higher-dimensional theories and those of lower dimensions and vice versa.
In what follows we explore two examples to demonstrate this procedure. In the rest of this section, we will consider static spherical solutions with the following metric ansatz
\begin{equation}  \label{s-ansatz}
d\hat{s}^2=-A dt^2+\frac{dr^2}{B}+\psi^2 (d\theta^2+ \sin^2 \theta d \varphi^2), \quad \psi(r)=r,
\end{equation}
and find the 2D theories and their corresponding solutions.

\subsection*{Example 1}
Consider the following model in 4D Horndeski gravity \eqref{Horndeski}
\begin{gather} \label{eq:Lag-EX_1-4D}
     \Lmh=(1+\beta \sqrt{-\Xh})\Rh-2 \Lambda+ \eta \Xh  -\frac{\beta}{2 \sqrt{- \Xh}}((\hat{\Box} \phi)^2-(\nabla_c \nabla_d \phi)^2),
 \end{gather}
where $G_2=\eta \Xh -2 \Lambda$, $G_4=1+\beta \sqrt{-\Xh}$,  $G_3=G_5=0$. In addition,
 $\eta$ and $\beta$ are constants and one may consider $\Lambda$ as cosmological constant. This theory admits a spherically symmetric  solution with the metric functions \eqref{s-ansatz} \cite{Babichev:2017guv}: 
\begin{gather} \label{sol1}
       A=B=1-\frac{2 m}{r}-\frac{\beta^2}{2 \eta r^2}-\frac{\Lambda r^2}{3}, \quad \phi(r)=\frac{\sqrt{2}\beta}{\eta}\int_{r}  \frac{\dd{r'}}{  \sqrt{A} \,r'^2}, \qquad \frac{\beta}{\eta}>0.
\end{gather}
The metric of this space-time is the same as the (Anti-)de Sitter-Reissner–Nordstr\"{o}m black hole and hence admits an event horizon. It is worth mentioning that even though $\phi^{\prime}(r)$ is singular on the horizon of the solution, $\phi(r)$ is smooth and well-defined. The thermodynamics of this black hole has been studied in \cite{Hajian:2020dcq}. 
Applying  the KK reduction on 2D  sphere using the metric ansatz \eqref{s-ansatz} to the Lagrangian \eqref{eq:Lag-EX_1-4D} leads to a two-dimensional theory \eqref{eq:2Dbiscalar} 
\begin{equation}
\begin{gathered}  \label{eq:Lag-EX_1-2D}
   F_2=\bx^2 (\eta \XX[11] -2 \Lambda)+2(1+2\XX[22])\psi^{-2} F_4 -4\beta \frac{\XX[12]^2}{\sqrt{\XX[11]}} ,\\
F_3^1= 4\beta  \frac{\XX[12]\, \bx  }{\sqrt{-\XX[11]}} , \quad
F_3^2=-4 \psi^{-1} F_4, \quad
 F_4 =\psi^2\qty(1+\beta \sqrt{-\XX[11]}) ,
\end{gathered}
\end{equation}
or equivalently as \eqref{eq:2D-KGB-like} with the functions
\begin{equation}
\begin{gathered}
    f_{3}^{1} = 4 \psi \frac{\XX[12]}{\XX[11]} \qty(1+\beta \sqrt{-\XX[11]}), \quad
    f_{3}^{2} = 2 \psi \qty(2 \beta \sqrt{-\XX[11]}+\ln \XX[11]),\\
    f_{2} =-4 \XX[22] \qty(2 \beta \sqrt{-\XX[11]}+\ln \XX[11]),\\
    K^{\mu} = \frac{\psi^2}{\XX[11]}(1+\beta \sqrt{-\XX[11]}) W_{11}^{\mu}-2 \psi \qty(2 \beta \sqrt{-\XX[11]}+\ln \XX[11]) \nabla^\mu \psi.
    \end{gathered}
\end{equation}
It is easy to check that this theory admits following the solution 
 \begin{gather}
    ds^2=-A dt^2+\frac{dr^2}{A},\label{2d-ansatz}
\end{gather}
in which metric function $A$ and  scalar field $\phi(r)$ are given in \eqref{sol1} and 
the other scalar field is
$\psi(r)=r$. 
\subsection*{Example 2}
Consider 4D theory
 \begin{equation}
    \Lmh=\Rh-2\Lambda+4\alpha \Xh -2\gamma \phi \Gh^{cd}\nabla_{c}\nabla_{d}\phih ,
 \end{equation}
with $G_{2}=4\alpha\Xh-2\Lambda$, $G_3=0$, $G_4=1$, and $G_5=-2\gamma \phi$.
 As discussed in \cite{Rinaldi:2012vy}, this theory admits a spherical symmetric black hole solution \eqref{s-ansatz} where 
\begin{gather}
    A=1-\frac{2m}{r}+\frac{\alpha(4 \alpha-\lambda)}{3 \gamma(4 \alpha+\lambda)}r^2+\frac{\lambda^2\sqrt{\frac{\gamma}{\alpha}}}{(16 \alpha^2-\lambda^2)r} \tan^{-1}(\frac{r}{\sqrt{\frac{\gamma}{\alpha}}}),\nonumber\\
    B=\frac{(\gamma+\alpha r^2)A}{\gamma(r A)^{\prime}}, \quad \phi(r)=\int_{r}\sqrt{\frac{-\lambda (r^2 A^2)^{\prime}r}{8(\gamma+\alpha r^2)^2 A^2}} \dd{r}, \label{sol2}
\end{gather}
and $\lambda=2(\alpha+\gamma \Lambda)$. The thermodynamics of this black hole has been studied in \cite{Hajian:2020dcq}. Reducing the dimension on the sphere \eqref{s-ansatz}, will lead to a bi-scalar theory \eqref{eq:2Dbiscalar}
\begin{align} \label{eq:Lag-EX_2-2D}  
    F_2&=\bx^2 (4\alpha\XX[11]-2\Lambda)+2(1+2\XX[22]),\quad
F_3^1=2\gamma \phi(1+2\XX[22]) , \\
F_3^2&=-4 \qty( \bx -2\gamma \phi\XX[12]), \quad
 F_4 =\psi (\psi -4\gamma \phi \XX[12]),\nonumber
\end{align}
or equivalently as \eqref{eq:2D-KGB-like} with the functions
\begin{equation}
\begin{gathered}
    f_{3}^{1} =-4 \qty(\gamma (\phi \XX[22]+2 \psi \XX[12])-\psi \frac{\XX[12]}{\XX[11]}), \quad
    f_{3}^{2} = -2 \qty(2 \gamma(2 \phi \XX[12]+\psi \XX[11])-\psi \ln \XX[11], )\\
    f_{2} =-4 \qty(-4 \gamma \XX[12]^2+\XX[22] \ln \XX[11]),\\
    K^{\mu} = \frac{\psi^2}{\XX[11]} W_{11}^{\mu}-4 \gamma \phi \psi W_{12}^{\mu}+4 \gamma \XX[12](\phi \nabla^\mu \psi+\psi \nabla^\mu \phi)-2 \psi \ln \XX[11] \nabla^\mu \psi.
    \end{gathered}
\end{equation}
One can check that the functions \eqref{sol2} and $\psi(r)=r$ with ansatz \eqref{2d-ansatz}, are a solution to this 2D theory.
\section{Perturbation and propagating mode}\label{sec:pert}
As stated in the introduction, one of the aims of this manuscript is to establish a two-dimensional scalar-tensor gravity theory with a propagating mode. Therefore, in this section, we will examine linear perturbations around a general static solution of bi-scalar theory
to demonstrate that it allows for the propagation of a scalar mode. As the explicit expressions are too complicated and messy, we will only explain the steps and general structures of the equations. It's worth noting that a similar study has also been done for the 4D Horndeski theory in \cite{Kobayashi:2012kh, Kobayashi:2014wsa,Kobayashi:2011nu,Mironov:2024idn}.

Let us consider a static solution to the bi-scalar theory. Thanks to the simplicity of two dimensions, one can always use diffeomorphisms to bring it into the following form 
\begin{equation} \label{eq:bacground-metric}
g^{(0)}_{\mu \nu}dx^{\mu}dx^{\nu}=-A dt^2+\frac{dr^2}{B}, \quad \phi^{(0) I}=\phi^{I}(r).
\end{equation}
One may also define the $r$ coordinate to coincide with one of the scalars, specifically $\psi(r)=r$. We take this solution and perturb it:
\begin{align}
    g_{\mu \nu}=g^{(0)}_{\mu \nu}(r)+\epsilon \; \delta g_{\mu \nu}(t,r), \qquad \phi^{I}(t,r)=\phi^{(0) I}+\epsilon \; \delta \phi^{I}(t,r),
\end{align}
where metric perturbations are defined as 
\begin{align}
    \delta g_{tt}=A \,
    \mathrm{h}_{0}(t,r),\qquad \delta g_{rr}=\frac{\mathrm{h}_1(t,r)}{B} 
   \qquad \delta g_{tr}=\mathrm{h}_2(t,r).
\end{align}
In principle, it is possible to use the gauge freedom $x^{\mu} \to x^{\mu}+\xi^{\mu}$ to eliminate two perturbations. However, we currently do not utilize this freedom because it enables us to explicitly see all constraint equations. Additionally, eliminating two perturbations does not fix all residual gauge freedoms. Now we can expand the action to the second order 
 in $\epsilon$
\begin{equation}
\sqrt{g}\mathcal{L}=\mathcal{L}^{(0)}+\epsilon^{1}\mathcal{L}^{(1)}+\epsilon^{2}\mathcal{L}^{(2)}+\cdots\,.
\end{equation}
The zeroth-order term describes the reduced action for the background fields $g^{(0)}_{\mu \nu}$ and $\phi^{I}_0$, while the linear term vanishes due to the background field equations. The second-order Lagrangian $\mathcal{L}^{(2)}$ governs the dynamics of perturbations over the fixed background \eqref{eq:bacground-metric}. For a general bi-scalar theory \eqref{eq:2Dbiscalar} or \eqref{eq:2D-KGB-like-2} and after performing several integration by parts over $\mathrm{h}_i$'s, $\mathcal{L}^{(2)}$ reduces to
\begin{equation}
\mathcal{L}^{(2)}=\mathrm{h}_{0}^2 \mathcal{A}_{0}+\mathrm{h}_{1}^2 \mathcal{A}_{1}+\mathrm{h}_{0} \mathcal{C}_{0}+\mathrm{h}_{1} \mathcal{C}_{1}+\Gamma.
\end{equation}
The background field equations for $g^{(0)}_{tt}$ and $g^{(0)}_{rr}$ indicate that $\mathcal{A}_{0}=0$ and $\mathcal{A}_{1}=0$ respectively. Therefore, there are no quadratic terms for $\mathrm{h}_0$ and $\mathrm{h}_1$ in $\mathcal{L}^{(2)}$. As a result, they do not appear in the field equations, and one should consider them as Lagrangian multipliers for the constraints
$\mathcal{C}_1(\dot{\delta \phi^{I}},\dot{{\delta \phi^{\prime I}}}, \dot{\mathrm{h}}_2)$=0, 
and $\mathcal{C}_0({\delta \phi^{\prime I}},{\delta \phi^{\prime \prime I}},\mathrm{h}_2^{\prime})=0$, where $'$ and $\dot{}$ indicate derivative with respect to $r$ and $t$ respectively. These two constraints are not independent and  consistent up to a possible constant and we can solve them to get $\mathrm{h}_2$ in terms of $\delta \phi^{I}$'s and their first derivative:
\begin{equation}\label{eq:h2}
    \mathrm{h}_2=a_I\delta \phi^{I}+b_I\delta \phi^{\prime I}.
\end{equation}
Where the explicit form of these coefficients is given in \eqref{eq:H2coef}.
In addition, $\Gamma$ depends on $\delta \phi^{I}$, $\mathrm{h}_2$ and their derivatives and by substituting  $\mathrm{h}_2$ we obtain an action for $\delta \phi^{I}$. After performing some integration by parts and adding a total derivative to remove terms like $\phi^{I}\partial_{t}\partial_{r}\phi^{I}$, we can express it in the following conical form:
\begin{equation}\label{eq:Lag-pert}
\mathcal{L}^{(2)}=\eval{\Gamma}_{\mathrm{h}_{2}}=\mathcal{T}_{IJ}\delta \dot{\phi^{I}}\dot{\delta \phi^{J}}-\mathcal{R}_{IJ}\delta{ \phi'}^{I}\delta \phi'^{J}+\mathcal{M}_{IJ}\delta \phi^{I}\delta \phi^{J}+\mathcal{N}_{IJ}\delta \phi^{I}\delta \phi'^{J}.
\end{equation}
In the $\delta \phi^{I}$ basis, the matrices are not diagonal, and there is a mixing between $\delta \phi^{1}$ and $\delta \phi^{2}$, hence to learn about the propagating modes, we need to analyze $\mathcal{T}_{IJ}$ and $\mathcal{R}_{IJ}$.  Specifically,  the stability (no-ghost condition) necessitates that all eigenvalues of $\mathcal{T}_{IJ}$, $\mathcal{R}_{IJ}$ be non-negative. It should be noted that in general, $\mathcal{M}_{IJ}$  and $\mathcal{N}_{IJ}$ might also provide some further stability conditions. However for simplicity, in the subsequent discussion, we focus on $\mathcal{T}_{IJ}$ and $\mathcal{R}_{IJ}$, arguing that these matrices are singular and possess zero eigenvalues. 

To understand this, we should note that there are two gauge freedoms associated with two possible linear coordinate transformations. We can use one of them to set $\mathrm{h}_{1}=0$. Now we can consider $\mathcal{L}^{(2)}$ and  field equation for $\mathrm{h}_2$, which establish a relationship among $(\mathrm{h}_{0},\mathrm{h}_{2},\delta \phi^{I})$. This relation along with \eqref{eq:h2} allows us to determine $\mathrm{h}_{0}$ as function  of other fields $(\delta \phi^{1},\delta \phi^{2})$. It should be noted that up to this point, we have utilized only one of the gauge freedoms to fix $\mathrm{h}_1$, whereas $\mathrm{h}_2$  and $\mathrm{h}_0$ are obtained from field equations (and residual gauge freedom). Consequently,  we are left with two degrees of freedom $(\delta \phi^{1},\delta \phi^{2})$, and another gauge transformation suggests that there is only one single propagating mode. This argument demonstrates that, in the appropriate gauge, $\mathcal{L}^{(2)}$ depends only on a single field. Now consider an appropriate basis where matrices are diagonal. Since there is only one mode in the system, the matrices should have just one non-zero component, which corresponds to the non-zero eigenvalue. We will verify this claim by computing them through simple examples.

Based on this argument, the stability condition for the propagating mode requires that the non-zero eigenvalue of $\mathcal{T}_{IJ}$  and $\mathcal{R}_{IJ}$ that we denote them by $\lambda_{\mathcal{T}}$ and
$\lambda_{\mathcal{R}}$ be positive.  Additionally, the propagation speed $c_{s}$  can be derived from the ratio of non-zero eigenvalues:
 \begin{equation}
     c_{s}^{2}=\frac{1}{A B}\frac{\lambda_{\mathcal{R}}}{\lambda_{\mathcal{T}}}.
 \end{equation}
Where the $AB$ factor appears since the genuine time and distance are given by $dt \sqrt{A}$ and $dr /\sqrt{B}$, respectively. The propagation speed can be compared to that in the 4D Horndeski theory. Indeed, the 4D theory allows for two propagating modes in the radial direction that could be interpreted as the propagation of gravitational and scalar waves \cite{Kobayashi:2014wsa}.  What we have obtained here is consistent with the speed of the scalar wave. This is expected because the 2D bi-scalar emerges from a spherically symmetric Kaluza-Klein reduction which only allows for spherical symmetric s-wave mode $l=0$.

Before closing this section it is worthwhile to note that the existence of a zero eigenvalue implies that $\det\abs{\mathcal{T}_{IJ}}=0$, which indicates
\begin{equation}
  \mathcal{T}_{IJ}=  \mqty( \mathcal{T}_{11} & \qty(\mathcal{T}_{11} \mathcal{T}_{22})^{\frac{1}{2}} \\
 \qty(\mathcal{T}_{11} \mathcal{T}_{22})^{\frac{1}{2}}& \mathcal{T}_{22})
\end{equation}
and hence 
$\lambda_{\mathcal{T}}=\mathcal{T}_{11}+\mathcal{T}_{22}$. Additionally, this matrix can be diagonalized using the following basis:
\begin{equation}
    \delta \bar{\phi} =\frac{\qty(\mathcal{T}_{11} \mathcal{T}_{22})^{\frac{1}{2}}\delta \psi + \mathcal{T}_{11} \; \delta \phi }{\qty(\mathcal{T}_{11} \lambda_\mathcal{T})^{\frac{1}{2}}}, \qquad   \delta \bar{\psi} =\frac{\qty(\mathcal{T}_{11} \mathcal{T}_{22})^{\frac{1}{2}}\delta \psi - \mathcal{T}_{22} \; \delta \phi }{\qty(\mathcal{T}_{22} \lambda_\mathcal{T})^{\frac{1}{2}}}.
\end{equation}
where only one of them propagates. The similar argument is also valid for $\mathcal{R}_{IJ}$ and shows $\lambda_{\mathcal{R}}=\mathcal{R}_{11}+\mathcal{R}_{22}$. The explicit form of these coefficients is given in the Appendix \ref{sec:pert}. Now let us present the results of the above procedure for the examples in Sec. \ref{sec:bi-scalar solutions}.

\subsection*{Example 1}
Let us consider 2D theory \eqref{eq:Lag-EX_1-2D} that admits the static solution \eqref{sol1}. Perturbing this solution yields \add{to} an effective Lagrangian that governs the perturbations as described by \eqref{eq:Lag-pert}, in which
\begin{equation}
  \mathcal{T}_{IJ}=  \frac{\sqrt{2} \beta }{\sqrt{A}}\mqty(-\phi_{0}'^{\;-1} & 1 \\
 1 & -\phi_{0}'), \qquad \mathcal{R}_{IJ}= -\frac{A}{2}\mathcal{T}_{IJ}.
\end{equation}
The determinants of the matrix coefficients are zero, and the non-zero eigenvalues are given by
\begin{equation}
\lambda_{\mathcal{T}}=   \frac{-2 \beta ^2-\eta ^2 r^4 A}{\eta  r^2 A}, \qquad \lambda_{\mathcal{R}}=-\frac{A}{2}\lambda_{\mathcal{T}},
\end{equation}
where we have used \eqref{sol1} to simplify the relations. One may note that the no-ghost condition $\lambda_{\mathcal{T}}>0$ indicates that $A<0$. In addition,  we find that the perturbation propagates with the following speed
 \begin{equation}
     c_{s}^{2}=-\frac{1}{2A}.
 \end{equation}
 One may verify that this outcome aligns with the speed of the scalar wave in the 4D Horndeski theory. It is important to note that the stability condition $A<0$ implies that the corresponding 4D solutions, as given by \eqref{sol1}, can not describe a stable black hole if $t$ is considered a timelike coordinate. Additionally,  $\abs{A}<\frac{1}{2}$ results in superluminal mode.

 We can identify the propagating mode by transforming to a normal basis where $\mathcal{T}_{ij}$ and $\mathcal{R}_{ij}$ are diagonal 
\begin{equation}
    \delta \bar{\phi} =\frac{\delta \psi + \phi _0' \; \delta \phi }{\sqrt{1+\phi _0' {}^2}}, \qquad   \delta \bar{\psi} =\frac{\phi _0'\; \delta \psi - \; \delta \phi }{\sqrt{1+\phi _0' {}^2}} .
\end{equation}
This transformation demonstrates that only $\delta \bar{\psi}(t,r)$ represents a propagating perturbation.
\subsection*{Example 2} 
In the case of the second example, 2D theory \eqref{eq:Lag-EX_2-2D} with the background solution  \eqref{sol2}, the expressions are more complicated however  we find the asymptotic expression for $c$ as
\begin{equation}
 \lim_{r\to \infty} c_{s}^2= 1-\frac{4\alpha}{\alpha+3\gamma \Lambda}.
\end{equation}
The result reported in \cite{Kobayashi:2014wsa} for the corresponding 4D theory is identical to the one mentioned above. Note that stability and subliminal conditions restrict the parameter space of  $\alpha$, $\gamma$, and $\lambda$ in the theory presented in \eqref{eq:Lag-EX_2-2D}. For example, for $\Lambda<0$ and $\alpha<0$ the subliminal and stability conditions require that $\gamma>\alpha/\Lambda$.

\section{Single-scalar theories} \label{sec:signle-scalar}
Having a bi-scalar theory in hand, there are several ways to restrict it to a single-scalar theory. For instance, one can investigate the case in which the scalar fields in Lagrangian \eqref{eq:2Dbiscalar} are the same: $\bx(x)=\phi(x)$. Then the Lagrangian reduces to 
\begin{align}\label{2DLagrangian}
 \hlbox{ \Lm_{\mathrm{H}}= F_2 + F_{3}  \Box \phi + F_4 R + F_{4,\X}\qty((\Box \phi)^2 -(\nabla_{\mu}\nabla_{\nu}\phi)^2) ,}
\end{align}
in which $F_n$ is a function of $\phi$ and $\X=\XX[11]$
\begin{align}\label{2DLagrangianFunctions}
  F_2(\phi,\X)&= \phi^2 G_2 - 4 \phi \X G_3 + 2 (1+ 2 \X) G_4 + 8  \X^2 G_{4,\X} \; ,\nonumber \\
  F_3(\phi,\X)&= \phi^2 G_3 -4 \phi G_4 -8 \phi \X G_{4,\X}-(1+6 \X) G_5 -4\X^2 G_{5,\X} \;,\\
  F_4(\phi,\X)&=\phi^2 G_4 + 2 \phi \X G_5 \; .\nonumber
\end{align}
Even though the reduced Lagrangian \eqref{2DLagrangian} has a specific relation to the higher dimensional functions \eqref{2DLagrangianFunctions}, as far as $G_n$ are arbitrary functions, $F_n$ are also general. Therefore, one may assume $F_n$ are arbitrary functions of  $\phi$ and $\X$. 

Furthermore, it's worth noting that this Lagrangian is identical to the four-dimensional Horndeski theory \eqref{Horndeski} up to the $\Lmh_{5}$ family that vanishes identically in two dimensions. However, due to identity \eqref{identity:single-scalar} for the special case $\phi=\psi$, it can be rewritten in another form.
\begin{gather} \label{Lag:2D Horndeski}
    \Lm_{\mathrm{H}}'= H_2(\phi,\X) + H_{3} (\phi,\X) \Box \phi \;=\Lm_{\mathrm{H}}+\nabla_{\mu}W^{\mu}.
\end{gather}
where the explicit form of $W^{\mu}$ is given in \eqref{identity:single-scalar}. This theory is the single scalar field version of \eqref{eq:2D-KGB-like-2}.
Indeed, the recent Lagrangian has been introduced by Horndeski in \cite{Horndeski:1974wa} as the most general two-dimensional scalar-tensor theory with second-order field equations. Interestingly, as has been argued in \cite{Nejati:2023hpe}, the Horndeski theory $\Lm_\mathrm{H}$ and $\Lm_\mathrm{H}'$ can be generated from JT Lagrangian \eqref{JT} through an invertible disformal transformation $ g_{\mu \nu } \to \mathcal{A} (\phi,\X) g_{\mu \nu}+\mathcal{B} (\phi,\X) \nabla_\mu \phi \nabla_\nu \phi$.

In the rest of this section, we will explore special cases of 2D Lagrangian given by equation \eqref{2DLagrangian} for functions of \eqref{2DLagrangianFunctions}. In particular, let us assume $\F_4=\ff(\phi)$ and $\F_3=\ff_3(\phi)$. The first assumption vanishes the term proportional to $\F_{4,\X}$ from the Lagrangian, and the second assumption allows us to combine $\F_2$ and $\F_3$ 
$$V(\phi,\X):=\F_2+\F_3 \Box \phi=\F_2+2\X \,\ff_{3,\phi}+\nabla_{\mu}(\ff_{3} \nabla^{\mu}\phi)$$ 
to obtain generalized dilaton gravity
\begin{gather}\label{GRZ}
    \Lm_{\text{GRZ}}=\ff(\phi) R +  V(\phi,\X),
\end{gather}
up to a boundary term proportional to $\ff_{3} \nabla^{\mu}\phi$.
Note that this Lagrangian for $\ff(\phi) = \phi$ and $V=-2\mathcal{V}$ is the same as the original GRZ theory \cite{Grumiller:2021cwg}. This theory represents the most consistent Lorentz invariant deformation of JT gravity, which emerges from the gauge theoretic formulation of dilaton gravity as a Poisson sigma model. It is worth mentioning that 2D models have been explored in \cite{Almheiri:2014cka} and \cite{Witten:2020ert} are specific cases of the GRZ.

Now, if we further restrict the potential as  $V=-8\X+4\lambda^2 \phi^2$ and $\ff=\phi^2$ we get CGHS theory \cite{Callan:1992rs, Strominger:1994tn}
\begin{gather}\label{CGHS}
\Lm_{\text{CGHS}}=\phi^2 R-8\X+4\lambda^{2}\phi^2,
\end{gather}
{which is a renormalizable theory of quantum gravity that can be exactly solvable classically}. Eventually, the JT gravity is a special case where $V=2 \lambda \phi$ and $\ff=\phi$
\begin{gather}\label{JT}
    \Lm_{\text{JT}}=\phi \qty(R+ 2 \lambda).
\end{gather}
With a clear map established between the 4D Horndeski theory and 2D single-scalar theory, we can explore higher dimensional embedding of  GRZ, CGHS, and JT within the context of the 4D Horndeski theory. In the following, we will explicitly classify all 4D Horndeski theories that permit these specific theories.

Let's substitute the first assumption $\F_4=\ff(\phi)$ into \eqref{2DLagrangianFunctions} to get
\begin{gather} \label{eq:Xindependance}
    \phi \,\GG_{4,\X}+2 \partial_{\X} (\X \GG_5)=0.
\end{gather}
This equation fixes $G_5$ in terms of 
$G_4$ up to an arbitrary function $c(\phi)$
\begin{equation}
     G_{5}=\frac{-1}{2\X}(\phi G_{4}+c(\phi)).
\end{equation}
To have $\F_4=\ff(\phi)$,
we take $c=-\ff/\phi$.
Now, the assumption $\F_3=\ff_3(\phi)$ restricts $G_3$ as
\begin{equation}
    G_3= \frac{\ff_3}{\phi^2}+\frac{(1+2\X)\ff}{2\phi^3 \X}-\frac{(1-6\X)}{2 \phi \X} G_4+\frac{6  \X G_{4,\X}}{\phi}.
\end{equation}
Utilizing this relation and definition of $V(\phi,\X)$ we get $G_2$
\begin{equation}
    G_2=\frac{V}{\phi^2}+\frac{2(1+2 \X)\ff}{\phi^4}-2 \X \partial_{\phi}\qty(\ff_3/\phi^2)-\frac{4(1-2 \X)G_4}{\phi^2}+\frac{16 \X^2 G_{4,\X}}{\phi^2}.
\end{equation}
Using these equations we can uplift a generic 2D GRZ theory into a 4D Horndeski with an arbitrary function $G_4(\phi,\Xh)$.  

\subsection*{Simple embedding of GRZ}
Let's consider an illustrative example by setting
\begin{equation}
\begin{gathered}\label{eq:simple-model}
G_2=\frac{V}{\phi^2}+\frac{2(6\X-1)\ff}{\phi^4}{+8 \X \partial_{\phi}(\ff/\phi^3)}\\
G_3=0, \quad G_{4}=\frac{\ff}{\phi^2}, \quad G_5=0,
\end{gathered}
\end{equation}
where for simplicity we used $G_3\hat{\Box}\phi=2\Xh \,G_{3,\phi}+\nabla_{c}(G_{3} \nabla^{c}\phi)$ to eliminate $G_3$ in terms of $G_2$, also assumed $\ff_3=0$. By construction, this model theory admits the GRZ as well as JT and CGHS for a suitable choice of $V$. Note that as $G_4\neq1$, we have a 4D scalar-tensor theory with non-minimal coupling interaction. However, It's worth noting for specific case $\ff(\phi)=\phi^2$, $G_4=1$ and hence the 4D theory is Einstein theory with  scalar field described by
  $G_2 =\phi^{-2}(V+4\Xh-2)$. Therefore, the CGHS model \eqref{CGHS} emerges from a simple 4D scalar-tensor theory belonging to the kinetic gravity braiding (KGB) model
  \begin{equation}
      \Lmh= \Rh-\frac{4\X+2}{\phi^2}+4\lambda^2.
  \end{equation}
\subsection{Consistency of Field equations}\label{sec:field-eq}
The KK reduction is an on-shell relationship between theories of different dimensions, where higher-dimensional theories can be reduced to lower-dimensional ones. In this sense, one should substitute the reduction ansatz into field equations. 
Since we have only explored this process at the action level, it is important to confirm that the reduced action produces the correct field equations. In what follows we check this consistency for the mentioned example by comparing the reduced 4D field equation and the 2D field equations deduced from the 2D action \eqref{2DLagrangian}.
Let's consider 4D metric field equations $\hat{\E}^{cd}=0$ for the theory \eqref{eq:simple-model} with $G_5=G_3=0$ and $G_{4,X}=0$
\begin{equation}\label{eom4Dg}
    \begin{aligned}
\hat{\E}^{cd}&=G_4 \LGh^{cd}+\frac{1}{2}\qty(2 G_{4,\phi} \hat{\Box}\phi-4\Xh G_{4,\phi\phi}-G_2)\gh^{cd}\\
&-\frac{1}{2}\qty(2G_{4,\phi\phi}+G_{2,\X})\nabla^c \phi \nabla^d \phi-G_{4,\phi}\nabla^c \nabla^d\phi.
\end{aligned}
\end{equation}
Besides, varying the action for this model with respect to the  scalar field gives the scalar field equation as
$\hat{\E}_{\phi}=-\frac{1}{\Xh} \nabla_{c}\hat{\E}^{cd}\nabla_{d}\phi=0$. 

As we are interested in the single scalar case we assume $\psi=\phi$ in the reduction ansatz \eqref{eq:reduction-ansatz}. Now by substituting this 
in the field equation $\hat{\E}^{cd}$ and represent its relevant components
    ${\E}^{\mu \nu}$
    in terms of 2D quantities we obtain the 2D field equation 
\begin{equation}
    \begin{aligned}
        \E^{\mu \nu}&=-\qty(\frac{\phi^2 G_2+2G_4}{2\phi^2}+\frac{\X}{\phi^3}\qty((G_4 \phi^2)_{,\phi}+3\phi^2 G_{4,\phi}+2\phi^{3}G_{4,\phi\phi})-\frac{(G_4 \phi^2)_{,\phi}}{\phi^2}\Box \phi)g^{\mu \nu}\\
        &-(\frac{G_{2,X}}{2}+G_{4,\phi \phi})\nabla^{\mu} \phi \nabla^{\nu} \phi-\frac{1}{\phi^2}{(G_4 \phi^2)_{,\phi}}\nabla^{\mu}\nabla^{\nu}\phi=0
    \end{aligned}
\end{equation}

For \eqref{eq:simple-model} this equation can be expressed as
\begin{equation}
    \begin{aligned}
        \phi^2\E^{\mu \nu}&=\frac{1}{2}\qty(2\ff'\Box \phi-4\X \ff''-V)g^{\mu \nu}
        -\frac{1}{2}(2\ff''+V_{,\X})\nabla^{\mu} \phi \nabla^{\nu} \phi-{\ff'}\nabla^{\mu}\nabla^{\nu}\phi
    \end{aligned}
\end{equation}
that is exactly the field equations of GRZ theory which are obtained from the variation of \eqref{GRZ} with respect to the metric (as one can compare it with \eqref{eom4Dg} for $G_2 \to V$ and $G_4 \to \ff$ and noting that 2D Einstein tensor is identical to zero).

Now we consider the field equation for the scalar field as
\begin{align}
\hat{\E}_{\phi}&=G_{4,\phi}\Rh+G_{2,\phi}+\nabla_{c}(G_{2,\X} \nabla^{c} \phi)\nonumber\\
&=G_{4,\phi}(R+ \phi^{-2}\Rm-4 \phi^{-1} \Box \phi +4 \phi^{-2}\X \; )+G_{2,\phi}+\nabla_{c}(G_{2,\X} \nabla^{c} \phi)=0,
\end{align}
where in the second line we use the reduction ansatz \eqref{eq:reduction-ansatz} for 
$\psi=\phi$. On the other hand the scalar field equation for the GRZ theory
\eqref{GRZ} is given by
\begin{equation}
    \E_{\phi}=\ff_{,\phi}R +V_{,\phi}+\nabla_{\mu}(V_{\X} \nabla^{\mu} \phi)=0
\end{equation}
It is easy to show that these two equations are related
\begin{align}
     \phi^2  \hat{\E}_{\phi}-\E_{\phi}&=(\phi^2 G_{4,\phi}-\ff_{\phi})R+(2-4 \phi \Box \phi+4\X)G_{4,\phi}\nonumber\\
     & +\phi^2 G_{2,\phi}-V_{,\phi}+\phi^{2}\nabla_{c}(G_{2,\X}\nabla^{c}\phi)-\nabla_{\mu}(V_{\X}\nabla^{\mu}\phi).
\end{align}
After using \eqref{eq:simple-model},  we might anticipate that the right-hand side of this equation should exhibit proportionality to the two-sphere component of the 4D equations, denoted as $\hat{\E}^{ij}$ --- as this part contributes to field equation $\psi$ through varying action with respect to $\gh^{ij}$ and we impose $\psi=\phi$ constraint. Intriguingly, a straight-forward calculation shows 
\begin{equation}
   \phi^2  \hat{\E}_{\phi}-\E_{\phi}=2\phi \E^{ij}\gh_{ij}.
\end{equation}
Therefore, the validity of 4D equations
$\hat{\E}_{\phi}=0$ and $\hat{\E}^{ij}=0$ guarantees the scalar field equation 
$\E_{\phi}=0$ that comes from GRZ action.

It is important to note that not every solution to the 2D theory can be uplifted to 4D. This can be seen by observing that the 2D field equation $\E_{\phi}=0$ only ensures that a specific combination of the 4D theory is satisfied $ \phi^2  \hat{\E}_{\phi}-2\phi \hat{\E}^{ij}\gh_{ij}=0$.
\section{Concluding Remarks}\label{sec:Remarks}
In this paper, we have investigated 2D bi-scalar theories of gravity through KK reduction of the 4D Horndeski theory over a 2D sphere. The radius of this sphere appears as an additional scalar field alongside the original Horndeski scalar field. The resulting theory \eqref{eq:2Dbiscalar} is similar to a generic 4D bi-scalar gravity \cite{Ohashi:2015fma, Horndeski:2024hee} up to terms that identically vanish in two dimensions (see appendix \ref{app:4D-bi-scalar}). 
Particularly in 2D bi-scalar theory, we discovered an interesting identity expressing the fourth family in terms of other families, excluding a total derivative that does not impact the classical field equations \eqref{eq:2D-identiy-L4-main}. Observing the simplified form of the bi-scalar theory, we conjectured the general form of the 2D bi-scalar theory. This proposal aligns with what is indicated by the generic form of 4D bi-scalar \ref{app:4D-bi-scalar}. 

Noting that KK reduction is an on-shell procedure, we mapped solutions of 4D  theories to 2D ones.
We discussed two examples of this approach by choosing  4D Lagrangians and their spherical symmetric solutions and identified the associated 2D theories and solutions. However, the inverse process is not straightforward, as it is not possible to uplift every  2D solution to its corresponding 4D theory.

We also examined linear perturbations over a generic static background solution in 2D bi-scalar theory.  Our analysis demonstrated that in this 2D theory, there is only one mode of propagation, and its speed is identical to that of a scalar wave in the 4D Horndeski theory.

With a bi-scalar theory in hand, we investigated a specific scenario where the two scalar fields are identical. This enabled us to obtain a generic single-scalar theory akin to the 4D Horndeski theory, except for the fifth family, which vanishes in two dimensions. Once again, using a two-dimensional identity, we expressed it in an alternative form, as initially introduced by Horndeski himself.  Note that this theory is the most general 2D single-scalar theory that admits second-order field equations. Furthermore, as argued in \cite{Nejati:2023hpe}, it involves generalized 2D dilaton gravity \cite{Grumiller:2021cwg} that encompasses JT and CGHS theories. Thus, our analysis enabled us to classify all possible embedding of these 2D theories into the 4D Horndeski theory. In this sense, adding extra scalar matter to 2D theory creates a certain bi-scalar $\psi\neq\phi$ that may uplifted to 4D theory. In addition, to verify the consistency of reduction, we showed that the field equations derived from the reduced 2D Lagrangian match those obtained from the reduction of higher-dimensional field equations.

Considering the result is fascinating for the study of thermodynamics in 4D Horndeski black holes through the 2D counterparts, which could provide deeper insights into the proposed modifications of the Hawking temperature as presented in \cite {Hajian:2020dcq}.  Notably, in 4D theory, the effective light cone linked to the gravity mode addresses the issue of the integrability of the black hole first law. Conversely, the 2D version of the theory suggests another effective light cone associated with the propagation of scalar waves. 

In this manuscript, we have focused only on the propagation of perturbation over a fixed background. However, it would be interesting to study the propagation of waves in fully non-linear regimes and figure out the causal structure according to \cite{Tanahashi:2017kgn, Reall:2021voz}.

Besides, it is intriguing to try to extend this work to construct multi-scalar 2D gravity by exploring KK reduction of higher dimension  Gallileons or Lovelock theories over $S^2\times \cdots\times S^2$. 

Finally, the KK reduction of more general 4D theories such as ``Degenerate Higher Order Scalar Tensor theories” (DHOST) \cite{Bettoni:2013diz} would be exciting. These theories may be obtained from generic disformal transformation to 4D Horndenski. 
On the other hand, the 2D Horndeski theory is invariant under disformal transformations \cite{Nejati:2023hpe, Takahashi:2018yzc}. In this sense, it is interesting to compare the disformal transformed 2D bi-scalar theory with the KK reduction of DHOST.

\section*{Acknowledgment}
We would like to express our gratitude to M.M. Sheikh-Jabbari for his invaluable comments and support, and to V. Taghiloo for the engaging discussions.  We are also thankful to D. Grumiller for his comment on the paper. MSN acknowledges the financial support received from OeaD (Ernst Mach worldwide).


\appendix
\section{More details on dimensional reduction} \label{app:KK}
In this section, we will present the details of dimensional reduction on a sphere from 4D to 2D. Decomposition of metric as \eqref{eq:reduction-ansatz}
\begin{eqnarray} 
  d\hat{s}^{2}&=&\gh_{cd} \; d\xh^{c} d\xh^{d}= g_{\mu \nu}(x) \; dx^{\mu} dx^{\nu}+\bx^{2}(x) \gamma_{i j}(\theta) \; d\theta^{i} d\theta^{j} ,
\end{eqnarray}
 will lead to $\sqrt{\abs{\gh}}=\psi^2 \sqrt{\abs{\gamma}}\sqrt{\abs{g}}$. Then 4D connections for 2D ones will be 
\begin{gather}
\hat{\Gamma}^{\mu}_{\nu \rho}=\Gamma^{\mu}_{\nu \rho}(g),\nonumber\\
 \hat{\Gamma}^{\mu}_{ij}=- \bx g^{\mu \nu} \gamma_{ij} \partial_{\nu} \bx, \nonumber\\
  \hat{\Gamma}^{j}_{i \mu} = \hat{\Gamma}^{j}_{ \mu i }=\bx^{-1}\delta^{j}_{i}\partial_{\mu}\bx ,\\
  \hat{\Gamma}^{k}_{ij}=\Gamma^{k}_{ij}(\gamma) ,\nonumber
\end{gather}
in which $\Gamma^{\mu}_{\nu \rho}(g)$ are 2D Christoffel symbols and $\Gamma^{k}_{ij}(\gamma)$ are the Christoffel symbols built on the 2-sphere. Then the Ricci tensor will be
\begin{gather}
    \Rh_{\mu \nu}={R}_{\mu \nu}-\frac{2}{\bx} \nabla_{\mu} \nabla_{\nu}\bx ,\\
     \Rh_{ij}={R}_{ij}-\gamma_{ij} \qty(\bx \Box \bx + \nabla_{\mu}\bx\nabla^{\mu}\bx).\nonumber
\end{gather}
Note that spherical symmetry indicates $\Rh_{\mu i}=\Rh_{i \mu}=0$. Otherwise, there will be a preferred direction that is not consistent with spherical symmetry. The 4D Ricci scalar concerning 2D ones is given as:
\begin{gather}
    \Rh=R+ \bx^{-2} \Rm-4 \bx^{-1} \Box \bx -2 \bx^{-2}\nabla_{\mu}\bx\nabla^{\mu}\bx \; .
\end{gather}
These results lead to Einstein tensors
\begin{gather}
  \LGh_{\mu \nu}=\cancelto{0}{G_{\mu \nu}}-\frac{2}{\bx}\nabla_{\mu}\nabla_{\nu}\bx-\frac{1}{2} g_{\mu \nu} \qty(\frac{\cancelto{2}{\Rm}}{\bx^2}-\frac{4}{\bx}\Box \bx -\frac{2}{\bx^2}\nabla_{\mu}\bx\nabla^{\mu}\bx)\nonumber\; , \\
  \LGh_{ij}=\cancelto{0}{\mathcal{G}_{ij}}+\gamma_{ij}\qty(\bx\Box \bx-\frac{\bx^2}{2} R)\; .
\end{gather}
Note that, even though the 2D Einstein tensor is identically zero, the 2D part of the 4D Einstein tensor is not zero. Besides, considering spherical symmetry leads to $\LGh_{\mu i}=0$. 4D derivatives of scalar field and 2D ones are related as
\begin{gather}
  \hat{\Box} \phi =\Box \phi + \frac{2}{\bx} \nabla_{\mu}\bx\nabla^{\mu}\phi \, ,\\
 (\nabla_{c}\nabla_{d}\phi)^ 2 =(\nabla_{\mu}\nabla_{\nu}\phi)^ 2+2 \bx^{-2}(\nabla_{\mu}\bx\nabla^{\mu}\phi)^2\; ,\\
  (\nabla_{c}\nabla_{d}\phi)^3=(\nabla_{\mu}\nabla_{\nu}\phi)^3+2 \bx^{-3} (\nabla_{\mu}\bx\nabla^{\mu}\phi)^{3} \; ,\\
   -\bx \LGh_{cd} \nabla^{c} \nabla^{d} \phih(\xh) =2 \nabla_{\mu}\nabla_{\nu}\bx \nabla^{\mu} \nabla^{\nu} \phi+\qty(\bx^{-1}-2  \Box \bx \X_{22}) \Box \phi+2 \X_{12}\qty( \frac{2\Box \bx}{\bx} -\mathrm{R}).
\end{gather}
Consequently, Kaluza-Klein reduction of Lagrangian (\ref{Horndeski}) leads to the following bi-scalar Lagrangian:
\begin{eqnarray} \label{bi-scalar}
	\Lm_{2} &=&  \bx^{2} \; G_{2},\nonumber\\
	\Lm_{3} &=& \bx^{2}  G_{3} \, \qty(\Box \phi + 2 \bx^{-1} \nabla_{\mu}\bx\nabla^{\mu}\phi) ,\nonumber\\
	\Lm_{4}&=&  \bx^{2} G_{4}  \, (R+2 \bx^{-2}-4 \bx^{-1} \Box \bx -2 \bx^{-2} \nabla_{\mu}\bx\nabla^{\mu}\bx)\qquad \\
	&+&  \bx^{2} G_{4,\mathrm{\X}}\qty((\Box \phi)^2-(\nabla_{\mu}\nabla_{\nu}\phi)^ 2+ 2 \bx^{-1} \nabla_{\mu}\bx\nabla^{\mu}\phi \qty(2\Box \phi+ \bx^{-1}\nabla_{\mu}\bx\nabla^{\mu}\phi)),\qquad\nonumber\\
	\Lm_{5} &=&\bx G_{5}\Big{(}-2 \nabla_{\mu}\nabla_{\nu}\bx \nabla^{\mu} \nabla^{\nu} \phi+\qty(-\bx^{-1}+2\Box \bx+\bx^{-1}\nabla_{\mu}\bx\nabla^{\mu}\bx) \Box \phi\\
 &+& \qty(2 \bx^{-1}\Box \bx -R) \nabla_{\mu}\bx\nabla^{\mu}\phi\Big{)}
	-\frac{1}{6} \bx^2   G_{5,\mathrm{\X}}
	\big{(}(\Box \phi)^{3} -3 \Box \phi (\nabla_{\mu} \nabla_{\nu} \phi)^2+ 2 (\nabla_{\mu} \nabla_{\nu} \phi)^3 \big{)} \nonumber\\
	&-& \bx  G_{5,\mathrm{\X}}
	\qty[ (\Box \phi)^2 -(\nabla_{\mu}\nabla_{\nu}\phi)^2+\bx^{-1}\Box \phi (\nabla_{\mu}\bx\nabla^{\mu}\phi)]  (\nabla_{\mu}\bx\nabla^{\mu}\phi),\qquad \qquad \nonumber
\end{eqnarray}
in which $G_{n}$ are functions of $\phi$ and $\X$, namely $G_n(\phi,\X)$. Note that in two-dimensional space-time, the following identity holds
\begin{gather}
    (\Box \phi)^{3} -3 \Box \phi (\nabla_{\mu} \nabla_{\nu} \phi)^2+ 2 (\nabla_{\mu} \nabla_{\nu} \phi)^3=0 .
\end{gather}
 As a result, the total reduced Lagrangian can be written as 
\begin{align}
\mathcal{L} &=\bx^2 G_2-4\bx \XX[12]G_3+2(1+2\XX[22])G_4 +8\XX[12]^2 G_{4,\XX[11]} \nonumber\\
&-4 \qty( \bx G_4 +\XX[12] G_5) \Box \bx+ (\bx^2 G_3-G_5 -8\bx \XX[12]  G_{4,\XX[11]}-2 G_5 \XX[22]-4\XX[12]^2 G_{5,\XX[11]} ) \Box \phi \nonumber\\
 &+(\psi^2 G_4+2 \psi G_5 \XX[12])R \\
&+\qty(\bx^2 G_{4, \XX[11]}+2\XX[12] \bx G_{5,\X})\qty(\Box \phi^2-(\nabla_{\mu}\nabla_{\nu}\phi)^ 2) +2 \psi G_5 (\Box \phi \Box \bx-\nabla_{\mu}\nabla_{\nu}\bx \nabla^{\mu} \nabla^{\nu} \phi) .\nonumber
\end{align}


    \section{Two-dimensional single-scalar identity} \label{app:id}
    Let's start from 2D identity \eqref{eq:2D-identiy-L4}
    \begin{equation}
         \X R +((\Box \phi)^2 -(\nabla_{\mu}\nabla_{\nu}\phi)^2)
  =\nabla_\mu \mathcal{W}^{\mu}, \quad  \mathcal{W}^{\mu}:=\nabla^\mu \X+ \Box \phi \nabla^\mu \phi,
    \end{equation}
    that arises from contraction 
     $\comm{\nabla_\mu}{\nabla_\nu} \nabla^\rho \phi= R^{\rho}_{\; \sigma \mu \nu} \nabla^\sigma \phi$ and 2D property of Ricci tensor $R_{\mu \nu}=\frac{R}{2}g_{\mu \nu}$. Now we multiply both sides by $F=F(\phi,\X)$ and use the Leibniz rule $F\nabla_{\mu}\mathcal{W}^{\mu}=\nabla_{\mu}(F \mathcal{W}^{\mu})-\mathcal{W}^{\mu}\nabla_{\mu}F$. Then by noting
     $H:=\X F$ and employing 2D identity 

 \begin{equation}
    \nabla_\mu \X \nabla^\mu \X +\Box \phi \nabla_\mu \X \nabla^\mu \phi= \X \qty((\Box \phi)^2-(\nabla_\mu \nabla_\nu \phi)^2).
\end{equation}
we obtain 
\cite{Takahashi:2018yzc,Nejati:2023hpe}
\begin{gather} \label{identity:single-scalar}
     H R + H_{,\X}((\Box \phi)^2 -(\nabla_{\mu}\nabla_{\nu}\phi)^2) 
  =-2 \X D_{,\phi}+\qty(2 H_{,\phi} +D)\Box \phi+\nabla_\mu W^\mu,\\
  W^\mu:=\frac{H}{\X}\mathcal{W}^{\mu}-D \nabla^\mu \phi, \quad D(\phi,\X):=\int_{\X} \frac{H_{,\phi}}{\X'} d\X'. \nonumber
 \end{gather}


\section{Two-dimensional bi-scalar identity}\label{app:identity-bi-scalar}
Having the identity \eqref{identity:single-scalar}, it is interesting to generalize it to the bi-scalar version. This is what we are going to present in what follows. 

Let us consider $M R+ M_{,\XX[IJ]} L_{IJ}$ that could be a generalization for the fourth part of bi-scalar theory \eqref{eq:2Dbiscalar}. Demanding this Lagraginan density admit second-order field equations, the function  $M(\phi^I, \XX[IJ])$ should obey the constraint  \cite{Ohashi:2015fma}
\begin{gather} \label{constraint}
M_{,\XX[12],\XX[12]}=4 M_{,\XX[11],\XX[22]}.
\end{gather}
It is possible to prove that the following identity 
\begin{eqnarray} \label{identity-general}
  M R+ M_{,\XX[IJ]} L_{IJ} &=&\nabla_\mu K^\mu+ m_{3}^{I} \Box \phi^{I}+m_2,
\end{eqnarray}
holds for certain functions $m_{n}$ as long as condition \eqref{constraint} is satisfied. 

To be more specific, let's focus on the case that is relevant to 2D bi-scalar theory \eqref{eq:2Dbiscalar}. So we assume 
$M=M(\XX[11],\XX[12])$ which $M$ does not depend on $\XX[22]$. Then 
 the constraint \eqref{constraint} restricts it to be linear in $\XX[12]$:
$ M=\XX[12] G(\XX[11])+H(\XX[11])$ as happens for $F_{4}$ in 2D bi-scalar theory \eqref{eq:2Dbiscalar}.
Then the identity
\begin{equation}
  M R+ M_{,\XX[11]} L_{11}+M_{,\XX[12]} L_{12} =\nabla_\mu \qty( \frac{HW_{11}^{\mu}}{\XX[11]} +\frac{\XX[12] W_{11}^{\mu}}{\XX[11]^2}   \I[\XX[11] G' ]+\frac{W_{12}^{\mu}}{\XX[11]}\I[G])   
\end{equation}
holds, where
\begin{equation}
    \begin{gathered}
W_{IJ}^{\mu}:=\nabla^{\mu}\XX[IJ]+\frac{1}{2}(\Box \phi^{I}\nabla^{\mu}\phi^{J}+\Box \phi^{J}\nabla^{\mu}\phi^{I}),
    \end{gathered}
\end{equation}
and $\I[F]=\int F d\X$. However, in the case of bi-scalar theory \eqref{eq:2Dbiscalar} the function $F_4$ also depends on scalar fields $\Phi^{I}$ therefore we generalized the function $M$ as  $M=\XX[12] G(\phi^I,\XX[11])+H(\phi^I,\XX[11])$. Now it is a matter of calculation to show that the identity can be generalized as
\begin{equation}\label{identity}
  { M R+ M_{\XX[11]} L_{11}+M_{\XX[12]} L_{12} =\nabla_\mu K^\mu+ f_{3}^{\phi} \Box \phi+ f_{3}^{\psi} \Box \psi+f_2 },
\end{equation}
where by using the notion 
$\X=\XX[11]$,  $\Y=\XX[22]$, ${\Z}=\XX[12]$, the boundary term $K^{\mu}$ and
functions $f_{3}^{I}$ are given by
\begin{subequations}\label{function-identity}
 \begin{align}
     f_2 &=-\frac{2}{\X} \I[ \Z \Y G_{,\psi ,\psi}+ 2\Z^2 G_{,\phi, \psi}]-2 \I[ \Z G_{,\phi, \phi}+\frac{\Y}{\X} H_{,\psi, \psi}+\frac{2 \Z }{\X}H_{,\phi, \psi}]-2\X \I[\frac{H_{,\phi,\phi}}{\X}]), \\
f_3^{\phi}&=\frac{\Y }{\X} \I[G_{,\psi}]+\frac{2 \Z^2 }{\X^2}\I[\X G_{,\psi ,\X}]+\I[\frac{H_{,\phi}}{\X}]+2( H_{,\phi}+\frac{ \Z}{\X} H_{,\psi}+\Z G_{,\phi}), \; \\
f_3^{\psi} &= \frac{2 \Z }{\X}\I[G_{,\psi}]+\I[G_{,\phi}+\frac{H_{,\psi}}{\X}],\\
K^\mu &=\frac{H X+\Z  \I[G_{,\X} \X]}{\X^2} W_{11}^\mu + \frac{\I[G]}{\X} W_{12}^{\mu} -(\frac{\Z }{\X} \I[G_{,\psi}]+\I[\frac{H_{,\psi}}{\X}]) \nabla^\mu \psi \nonumber\\ &-(\frac{\Z }{\X} \I[G_{,\phi}]+\I[\frac{H_{,\phi}}{\X}]) \nabla^\mu \phi.
 \end{align}   
 \end{subequations}
 It's worth noting that the mentioned identity implies the 
$F_4$ family in bi-scalar theory \eqref{eq:2Dbiscalar}
can be absorbed to $F_2$ and $F_3$ families up to a total derivative that does not contribute to the equation of motion.
\section{Comparison with 4D bi-scalar theory } \label{app:4D-bi-scalar}
As mentioned in Sec. \ref{sec:signle-scalar},  the structure of the 2D Horndeski theory is the same as the 4D one, up to the 2D identities. This suggests deriving the 2D bi-scalar Lagrangian by considering the 4D one and imposing the 2D identities.

The general 4D bi-scalar theory which leads to second-order equations of motion is described by the summation of the following Lagrangians \cite {Ohashi:2015fma}:
\begin{align} \label{4DBiscalar}
	\Lp_{1} &=\MG^{(1)}_{I}\hat{\Box} \phi^{I},\nonumber\\
	\Lp_{2}&=2 \MG^{(2)} R+ 2 \MG^{(2)}_{,IJ}\qty(\hat{\Box} \phi^{I}\hat{\Box} \phi^{J}-\nabla^d \nabla_c \phi^I \nabla^c \nabla_d \phi^J),\nonumber\\
	\Lp_{3}&= \MG^{(3)}_{IJK} \qty(\nabla_c \phi^I \nabla^c \phi^J \hat{\Box} \phi^K- \nabla_c \phi^I \nabla^d \phi^J \nabla^c\nabla_d \phi^K), \nonumber\\
	\Lp_{4}&=-4 \MG^{(4)}_{I} \Gh_{cd} \nabla^c \nabla^d \phi^I + \frac{2}{3} \MG^{(4)}_{I,JK} \big{(}\hat{\Box}\phi^{I} \hat{\Box}\phi^{J} \hat{\Box}\phi^{K}- \hat{\Box} \phi^I \nabla_c \nabla_d \phi^J \nabla^d \nabla^c \phi^K\nonumber\\
	& -2 \hat{\Box} \phi^K \nabla_c \nabla_d \phi^I \nabla^d \nabla^c \phi^J +2\nabla^d \nabla_c \phi^I \nabla^c \nabla_e \phi^J \nabla^e \nabla_d \phi^K\big{)}, \nonumber\\
	\Lp_{5} &= -4 \MG^{(5)}_{IJ} \Gh_{cd} \nabla^c \phi^I \nabla^d \phi^J \nonumber\\
	& +2 \MG^{(5)}_{IJ,KL}\big{(} \nabla_c \phi^I \nabla^c \phi^J \hat{\Box} \phi^K \hat{\Box} \phi^L - \nabla_c \phi^I \nabla^c\phi^J \nabla_f \nabla_e \phi^K \nabla^e \nabla^f \phi^L \nonumber\\
	& -2 \nabla_c \phi^I \nabla^d \phi^J \nabla^c \nabla_d \phi^K \hat{\Box} \phi^L +2\nabla_c \phi^I \nabla^d \phi^J \nabla^c \nabla^h \phi^K  \nabla_h \nabla_d \phi^L \big{)},\nonumber\\
	\Lp_{6}&=\MG^{(6)},\nonumber\\
	\Lp_{7}&=  2 \MG^{(7)}_{IJKLM} \qty( \nabla_c \phi^I \nabla^c \phi^J \nabla_e \phi^k \nabla^e \phi^L \hat{\Box} \phi^M -2 \nabla_c \phi^I \nabla^c \phi^J \nabla_e \phi^k \nabla^f \phi^L \nabla^e \nabla_f \phi^M),
\end{align}
where $\MG^{(1)}$, $\MG^{(2)}$, $\MG^{(3)}_{IJK}$ , $\MG^{(4)}_{I}$,$\MG^{(5)}_{IJ}$, $\MG^{(6)}$ and $\MG^{(7)}_{IJKLM}$ are arbitrary functions of $\phi^{I}$ and $X^{IJ}$ satisfying 
\begin{gather}
	\MG^{(2)}_{IJ,KL}=\MG^{(2)}_{IK,JL}, \quad \MG^{(3)}_{IJK} = \MG^{(3)}_{JIK}, \quad    \MG^{(5)}_{IJ} = \MG^{(5)}_{JI},\\
	\MG^{(7)}_{IJKLM} = -\MG^{(7)}_{KJILM}= -\MG^{(7)}_{ILKJM}= \MG^{(7)}_{JILKM},\nonumber
\end{gather}
where following \cite{Ohashi:2015fma}, the notation $A_{,IJ}:= \frac{1}{2}\qty(A_{,\XX[IJ]}+A_{,\XX[JI]})$ is used.

Investigating this Lagrangian in 2D\footnote{Considering the \eqref{4DBiscalar} Lagrangian in 2D means to consider all indices to be $\{0,1\}$.} shows that $\Lp_4,\Lp_5$ and $\Lp_7$ will be identical to zero and the 2D bi-scalar  Lagrangian will lead to:
\begin{align}
	\Lm_{\text{BS}} &= \MG^{(6)}+\MG^{(1)}_{I}\Box \phi^{I}+ 2 \MG^{(2)} R+ 2 \MG^{(2)}_{,IJ}\qty(\Box \phi^{I}\Box \phi^{J}-\nabla_\nu \nabla_\mu \phi^I \nabla^\mu \nabla^\nu \phi^J)\nonumber\\
	&+\MG^{(3)}_{IJK} \qty(\nabla_\mu \phi^I \nabla^\mu \phi^J \Box \phi^K- \nabla^\mu \phi^I \nabla^\nu \phi^J \nabla_\mu\nabla_\nu \phi^K).
\end{align}
It is easy to prove that the last parentheses of the Lagrangian can be rewritten in the forms of $\MG^{(1)}_{I}\Box \phi^{I}$, $\MG^{(6)}$ families and total derivatives. As total derivatives do not contribute to the equations of motion, we ignore them. The resulting two-dimensional bi-scalar Lagrangian will be:
\begin{align}	\Lm_{\text{BS}}&= \MG^{(6)}+\MG^{(1)}_{I}\Box \phi^{I}+ 2 \MG^{(2)} R+ 2 \MG^{(2)}_{,IJ}\qty(\Box \phi^{I}\Box \phi^{J}-\nabla_\nu \nabla_\mu \phi^I \nabla^\mu \nabla^\nu \phi^J).
\end{align}
To the equations of motion remain second order, it should respect the constraint $M^{(2)}_{,IJ,KL}=M^{(2)}_{,IK,JL}$ {and $\MG^{(1)}_{I,JK}=\MG^{(1)}_{J,IK}$}. 
This result is the same as \eqref{eq:2Dbiscalar} As we discussed in the appendix \ref{app:identity-bi-scalar}, considering the first constraint, { $\MG^{(2)}$} family can be absorbed into the other two families, up to a total derivative. Consequently, the final Lagrangian will be 
\begin{align}
	\Lm_{\text{BS}}&= \mathrm{\MG}^{(6)}+\mathrm{\MG}^{(1)}_{I}\Box \phi^{I},
\end{align}
which {is} consistent with the result \eqref{eq:2D-KGB-like-2}.

\section{Supplementary to section \ref{sec:pert}}
In what follows  $F_{,f g}$ indicates $\partial_{f}\partial_{g}F$ and $D(U):=(\ln{U})'=\dv{\ln U}{r}$.
\begin{equation}\label{eq:H2coef}
\begin{aligned}
c=&-2 B\qty(X(3 \psi ' \GGT_{{3,X}}^{\psi }+ \phi ' \GGT_{{3,X}}^{\phi })+ Y( \psi ' \GGT_{{3,Y}}^{\psi }+3 \phi ' \GGT_{{3,Y}}^{\phi })),\\
c\; b_1=&-4 B \qty(X \GGT_{{3,X}}^{\phi }+2 Z \GGT_{{3,X}}^{\psi }+Y \GGT_{{3,Y}}^{\phi }), \qquad b_2=\frac{2 - b_1 \phi'}{\psi'},\\
c\;a_1=&-2 B \Big{(} \phi ' \GGT_{{2,X}}^{{}}+\frac{1}{2} \psi ' \GGT_{{2,Z}}^{{}}- X D(A X) \GGT_{{3,X}}^{\phi } - Y D(AY) \GGT_{{3,Y}}^{\phi } -\frac{2 (A Z)' }{A}\GGT_{{3,X}}^{\psi } \\&+ \psi ' \GGT_{{3,\phi }}^{\psi }+\phi ' \GGT_{{3,\phi }}^{\phi }+ (\GGT_3^{\phi })'\Big{)},\\
c\; a_2=&-2 B \Big{(} \psi ' \GGT_{{2,Y}}^{{}}+\frac{1}{2} \phi ' \GGT_{{2,Z}}^{{}}-X  \GGT_{{3,X}}^{\psi }  D(AX) - Y \GGT_{{3,Y}}^{\psi } D(AY) -\frac{2 (A Z)'}{A}\GGT_{{3,Y}}^{\phi } \\& + \psi ' \GGT_{{3,\psi }}^{\psi }+\phi ' \GGT_{{3,\psi }}^{\phi }+(\GGT_3^{\psi })'\Big{)} ,
\end{aligned}
\end{equation}
\begin{equation}
    \begin{aligned}
    &2 (A B)^{1/2}\mathcal{T}_{11}  = 
     \GGT_{{2,X}}^{{}}+b_1 Y Y' \GGT_{{3,YY}}^{\phi }+2\GGT_{{3,\phi }}^{\phi }+\GGT_{{3,\phi X}}^{\phi } X( b_1 \phi '-2)+ b_1 Y \phi ' \GGT_{{3,\phi Y}}^{\phi }\\
     &+2 Y \GGT_{{3,\psi X}}^{\psi } \qty( b_1  \phi '-1)+X' \GGT_{{3,XX}}^{\phi } \qty(b_1 X+B \phi ')+\GGT_{{3,XY}}^{\phi } \qty(2 b_1 Z Z'+ B Y' \phi ')+ b_1 Y \psi '\\
     &+\GGT_{{3,\psi X}}^{\phi } \qty(b_1 X \psi '-2  Z) \GGT_{{3,\psi Y}}^{\phi }
     +\GGT_{{3,XZ}}^{\psi } Z' \qty(2b_1 Z+B \psi')+2\GGT_{{3,\phi X}}^{\psi } \qty( b_1 X \psi '- Z)\\
     &+Y \GGT_{{3,XY}}^{\psi } \qty(2b_1 Z D(\frac{Z^3}{X})+ B \psi ' D(Y))+ X\GGT_{{3,XX}}^{\psi } \qty(B \psi' D(X^2 Y)+b_1 Z D(X^3 Y))\\
     &+ Y \GGT_{{3,Y}}^{\phi } \qty(( b_1'-2 a_1)+\frac{b_1}{2} D(\frac{Y^2}{A B}))+X \GGT_{{3,X}}^{\phi } \qty( (b_1'-2 a_1 -\frac{2}{\phi'}D(X))+ \frac{b_1}{2} D(\frac{X^2}{A B}))\\
     &+2 Z \GGT_{{3,X}}^{\psi } \qty((b_1'-2 a_1)+\frac{b_1}{2} D(\frac{X Y}{A B}) -\frac{Y'}{\psi' Z}).
    \end{aligned}
\end{equation}
{\allowdisplaybreaks
    \begin{align*}
    &2(A B)^{1/2}\mathcal{T}_{22}= \GGT_{{2,Y}}^{{}}+2 \GGT_{{3,\psi }}^{\psi }+b_2 X (X' \GGT_{{3,XX}}^{\psi }+ \phi ' \GGT_{{3,\phi X}}^{\psi }+\psi ' \GGT_{{3,\psi X}}^{\psi })+\GGT_{{3,\phi Y}}^{\psi } ( b_2 Y \phi '-2 Z)\\
    &+2\GGT_{{3,\psi Y}}^{\phi } ( b_2 Y \phi '-Z)+\GGT_{{3,YZ}}^{\phi }  Z'(2 b_2 Z+ B \phi ')+\GGT_{{3,\psi Y}}^{\psi } Y(b_2 \psi '-2)+2\GGT_{{3,\phi Y}}^{\phi } X(b_2 \psi '-1)\\
    &+\GGT_{{3,XY}}^{\psi } \qty(2 b_2 Z Z'+ B \psi ' X')+\GGT_{{3,YY}}^{\psi } Y'\qty(b_2 Y +B \psi')+X \GGT_{{3,XY}}^{\phi } \qty(B D(X) \phi '+b_2 Z D(X^3 Y))\\
    &+ \GGT_{{3,YY}}^{\phi }Z \qty(2 b_2 Y D(\frac{Z^3}{X}) + B \psi' D(X Y^2))+X \GGT_{{3,X}}^{\psi } \qty(( b_2'-2 a_2)+\frac{1}{2}b_2 D(\frac{A X^2}{B^3}))\\
    &+ Y\GGT_{{3,Y}}^{\psi }  \qty(b_2'+\frac{b_2}{2} D(\frac{A Y^2}{B^3})-2 (a_2+\frac{D(Y)}{A \psi '}))+2 Z\GGT_{{3,Y}}^{\phi } \qty(b_2'- a_2-\frac{X'}{Z \phi }+\frac{ b_2}{2} D(\frac{A X Y}{B^3})).
    \end{align*}}
{\allowdisplaybreaks
    \begin{align*}
      &8 (A B)^{3/2}\mathcal{R}_{11}  = -4 A^2 B\Bigg{(}b_1^2 Y \qty(\GGT_{{2,Y}}^{{}}+Y \GGT_{{2,YY}}^{{}}+Z \GGT_{{3,\phi Y}}^{\psi }+Y \GGT_{{3,\psi Y}}^{\psi })+B \GGT_{{3,\phi }}^{\phi } (b_1 \phi'-2)\\
      & +\GGT_{{2,X}}^{{}} (b_1^2 X+B(b_1 \phi'-1))+X(b_1(2 B \phi '+b_1 X)-2 B)\qty(\GGT_{{2,XX}}^{{}}+ \GGT_{{3,\phi X}}^{\phi })\\
      &+Y \Big{(}\GGT_{{3,\psi X}}^{\psi } (b_1 (4 B \phi '+3 b_1 X) -2 B)+ b_1 \GGT_{{3,\phi Y}}^{\phi } (2 B\phi '+3 b_1 X)+2 b_1 \GGT_{{2,XY}}^{{}} (B \phi '+ b_1 X) \\
      &+ \GGT_{{2,ZZ}}^{{}} ( b_1^2 X+B(b_1 \phi '-\frac{1}{2}))\Big{)}+ b_1 B (\GGT_3^{\phi })'+\GGT_{{3,\phi X}}^{\psi } (4 b_1 B X \psi '+(3 b_1^2 X-2 B) Z)\\
      &+\GGT_{{2,XZ}}^{{}} (3 b_1 B X \psi '+2 (b_1^2 X-B )Z)+\GGT_{{3,\psi X}}^{\phi } ( 2 B ( b_1 X \psi '-Z)+ b_1^2 X Z)\\
      &+b_1 Y \Big{(}\GGT_{{3,\psi Y}}^{\phi } (2 B\psi '+3 b_1 Z)+\GGT_{{2,YZ}}^{{}} (B \psi '+2 b_1 Z)\Big{)}\Bigg{)}+2 A B \Bigg{(}2 A' B X^2 \GGT_{{3,XX}}^{\phi } (\phi'(\frac{B}{X}-\frac{b_1^2}{2}) +2 b_1 )\\
      &+ b_1 B Y^2 A'\GGT_{{3,YY}}^{\phi } (4-5 b_1 \phi ')- b_1 B\psi' (2A  \GGT_{{3,\phi }}^{\psi }+ b_1 Y^2 A' \GGT_{{3,YY}}^{\psi })- A b_1\GGT_{{2,Z}}^{{}} (B \psi '+2 b_1 Z)\\
      &+A' B X\GGT_{{3,XX}}^{\psi } \qty( \psi ' (6 B-5 b_1^2 X) +16 b_1 Z)\Bigg{)}+A^2 B^2 \Bigg{(}  b_1^2 \Big{(}3 \psi ' (\GGT_3^{\psi })'+3\phi ' (\GGT_3^{\phi })'+\frac{\GGT_2}{B}\\
      &-Y \psi' \GGT_{{3,Y}}^{\psi } D(A^5 Y) \Big{)}+\GGT_{{3,X}}^{\phi}  \Big{(}X b_1 (-b_1 \phi' D(A^5 X^3) +4 D(A^3 X)+4 B D(A) \phi' \Big{)}+8  b_1 Z\GGT_{{3,X}}^{\psi } D(A^3 )\\
      &-\frac{3}{8} b_1^2 \phi' D(A^5 X Z^2)-\frac{D(A)}{\phi'})-2 b_1 Y\GGT_{{3,Y}}^{\phi } (3 b_1 (\frac{1}{2}\phi'D(A^5 B Y^2)+1)-2 D(A^3 Y))\\
      &+4 Y Z\GGT_{{3,YZ}}^{\phi } \Big{(} b_1^2 \phi' D(\frac{A Z}{X})+\frac{b_1}{2} \qty((3-\frac{Y}{Z}) D(\frac{Z^2}{X})+  (D(A^4)-D(B)(1-\frac{Z}{X})))-\frac{1}{\phi'} D(A Y)\Big{)}\\
      &-4 Y \Big{(}\GGT_{{3,XY}}^{\psi } \psi ' \Big{(} B  D(Y)+ b_1^2 X D(\frac{A^3 Z^2}{Y^2})-2 b_1 Z D(\frac{A^2 Z}{Y^2}) \Big{)}+ X \GGT_{{3,XY}}^{\phi } \Big{(} \phi' b_1^2 D(\frac{Z^2}{Y^2 A^3})\\
      &+2b_1( D(\frac{B X^2}{A^4 Z^3})-1) +\frac{2}{\phi'} D(A^2 Y) \Big{)}+X \GGT_{{3,XZ}}^{\psi } \Big{(} b_1^2 \phi' D(\frac{A X}{Z})-b_1 D(\frac{A^2 X^2}{Z^3})+\frac{1}{\phi'} D(\frac{A X}{Z^2})\Big{)} \Bigg{)}
      \end{align*}}
{\allowdisplaybreaks
\begin{align*}
    &8 (A B)^{3/2}\mathcal{R}_{22}  = -4 A^2 b_2^2 B X \qty(\GGT_{{2,X}}^{{}}+X \GGT_{{2,XX}}^{{}}+X \GGT_{{3,\phi X}}^{\phi }+ Z \GGT_{{3,\psi X}}^{\phi })
    \\
    &-4 A B \Big{(} B b_2 \phi' \qty( A  \GGT_{{3,\psi }}^{\phi }+ \frac{b_2}{2} X^2 A' \GGT_{{3,XX}}^{\phi })+A b_2 \qty(\GGT_{{2,Z}}^{{}} (\frac{B}{2} \phi '+ b_2 Z)+ X \GGT_{{3,\phi X}}^{\psi } (2 B \phi '+3 b_2 Z))\Big{)}\\
    &-4 A^2 B \qty(b_2 X \GGT_{{2,XZ}}^{{}} (B \phi '+2 b_2 Z)+\GGT_{{3,\psi Y}}^{\phi } (4 b_2 B Y \phi '+3 b_2^2 Y Z-2 B Z))\\
    &-4 A^2 B \qty(\GGT_{{2,YZ}}^{{}} (3 b_2 B Y \phi '+2 b_2^2 Y Z -2  B Z)+ \GGT_{{3,\phi Y}}^{\psi } (2 b_2 B Y \phi '+ b_2^2 Y Z-2 B Z))\\
    &-4 A^2 B \Big{(}B \GGT_{{3,\psi }}^{\psi } (b_2 \psi '-2)+\GGT_{{2,Y}}^{{}} ( b_2 B \psi '+ b_2^2 Y- B)+X \GGT_{{3,\phi Y}}^{\phi } (4 b_2 B \psi '+3 b_2^2 Y-2 B)\\
    &+ X \GGT_{{3,\psi X}}^{\psi } (2 B \psi '+3 b_2 Y)+2 b_2 X \GGT_{{2,XY}}^{{}} ( B \psi '+ b_2 Y)+\frac{1}{2} b_2 X \GGT_{{2,ZZ}}^{{}} (B (2 \psi'-1)+2 b_2 Y )\\
    &+Y (\GGT_{{3,\psi Y}}^{\psi }+ \GGT_{{2,YY}}^{{}} ) (2 b_2 B \psi '+b_2^2 Y-2 B)\Big{)}+2 A B^2 A' Y \GGT_{{3,YY}}^{\psi } (b_2 Y(- b_2 \psi '+4)+2 B \psi ')\\
    &+2 A A' B^2 Y \qty( \GGT_{{3,YY}}^{\phi } \qty((-5 b_2^2 Y+6 B )\phi '+16 b_2 Z)+ b_2 X^2\GGT_{{3,XX}}^{\psi } (4-5 b_2 \psi '))\\
    &- A^2  B^2 b_2 \qty(4(\GGT_3^{\psi })'+ b_2 X \GGT_{{3,X}}^{\phi } D(A^5 X^3)-b_2 (3 \psi ' (\GGT_3^{\psi })'+3 \phi ' (\GGT_3^{\phi })'+\frac{\GGT_2}{B}))\\
    &+A^2 B^2 \Bigg{(} b_2 \GGT_{{3,X}}^{\psi } \qty(-3 b_2 Z \phi' D(\frac{A^5 Z^4}{Y})+4 X D(A^3 X))+\GGT_{{3,Y}}^{\phi } \Big{(}-3 Y b_2^2 \phi' D(A^5 Y Z^2)\\
    &+8 b_2 Z(D(A^3 Z))+4 B  \phi' D(A)\Big{)}+\GGT_{{3,Y}}^{\psi } \Big{(}4 (Y b_2 D(A^3 Y)+B D(A) \psi ')-b_2^2 Y \phi' \frac{Y}{Z} D(A^5 Y^3)\Big{)}\\
    &-4 \GGT_{{3,XY}}^{\phi } X \Big{(}B \phi' D(X) +\phi' Y b_2^2 D(\frac{A^3 Y^2}{Z^2})- b_2 A (4 Z D(A) +(B-2 Z) D(B)-B D(\frac{Z^3}{Y^2}))\Big{)}\Bigg{)}\\
    &+2 A B^2 X \Bigg{(}\GGT_{{3,XZ}}^{\psi } \Big{(}-2 \phi' A b_2^2 Y D(\frac{A Z}{Y})+2b_2 A Z D(\frac{A^2 Z^3}{Y^2})-\frac{2}{\phi'}(A X)'\Big{)}\\
    &+ A Y \Big{(}\GGT_{{3,YZ}}^{\phi }\Big{(} b_2 ( \psi' D(A^2 \frac{Y}{X})- b_2 D(\frac{A^4 Y}{X^3}))- \frac{B}{Y} \psi ' D(\frac{A}{X})\Big{)} \\
    &+2 \GGT_{{3,XY}}^{\psi } \Big{(} \psi' b_2^2 D(\frac{Z^2}{A^3 X^2})+b_2 D(\frac{A^8 X^3}{Y})+\frac{B}{Y} \psi ' D(A^2 X) \Big{)}\Big{)}\Bigg{)}
    \end{align*}}


\addcontentsline{toc}{section}{References}
\bibliographystyle{fullsort.bst}
\bibliography{Bibliography}

\end{document}